\documentclass[aps,prd,twocolumn,showpacs,10pt,superscriptaddress,preprintnumbers,nofootinbib]{revtex4-1}
\usepackage[utf8]{inputenc}
\usepackage{amsmath,amssymb,bm,slashed,braket}
\usepackage{graphicx}
\usepackage{float}
\usepackage[dvipsnames,table,svgnames]{xcolor}
\usepackage[normalem]{ulem}
\usepackage{colortbl}
\usepackage{multirow,boldline, booktabs}
\usepackage{makecell}
\usepackage{hhline}
\usepackage{dsfont}
\usepackage{verbatim}
\usepackage{MnSymbol}
\usepackage[colorlinks=true,
            linkcolor=blue,
            urlcolor=blue,
            citecolor=teal,          
            bookmarks=true,
            bookmarksnumbered=true,
            breaklinks=true,
            pdfpagemode=Fullscreen,
            pdfstartview=FitBH]{hyperref}
\allowdisplaybreaks[4]
\usepackage[capitalise]{cleveref}
\usepackage{orcidlink}
\usepackage{pifont}
\usepackage{cancel}
\usepackage{tikz} 
\usepackage{tkz-euclide}
\usetikzlibrary{backgrounds} 
\usetikzlibrary{decorations.pathmorphing}
\usetikzlibrary{arrows.meta}
\usetikzlibrary{shapes.misc}
\tikzset{
mystyle/.style={line width=1, baseline, scale=0.6, every node/.style={scale=1}},
photon/.style={decorate, decoration={snake, segment length=1.5 mm, amplitude=0.5mm}, draw=black, thick},
v/.style={decorate, draw, decoration={snake, segment length=2.mm, amplitude=0.5mm}},
f/.style={draw, decoration={markings,mark=at position #1 with {\arrow[]{Latex[length=1.5mm,width=1.7mm]}}},
    postaction={decorate},node contents=#1},
f/.default=.6,
fb/.style={draw,decoration={markings,mark=at position #1 with {\arrowreversed[]{Latex[length=1.5mm,width=1.5mm]}}},
    postaction={decorate},node contents=#1},
fb/.default=.6,
s/.style={dashed,draw, postaction={decorate},
        decoration={markings,mark=at position .55 with {\arrow[]{Latex[length=1.5mm,width=1.7mm]}}}},
sb/.style={dashed,draw, postaction={decorate},
        decoration={markings,mark=at position .55 with {\arrowreversed[draw=black,very thick]{latex}}}},
snar/.style={dashed,draw,line width =1.25pt},
gluon/.style={decorate,
 decoration={coil,amplitude=2pt, segment length=3.5pt,  pre length=.1cm, post length=.1cm}},
}

\newcommand{\calO}{\mathcal{O}}
\newcommand{\calP}{\mathcal{P}}
\newcommand{\calN}{\mathcal{N}}
\newcommand{\tN}{ \textsc{n} }
\newcommand{\C}{ \textsc{c} }
\newcommand{\tL}{ \textsc{l} }
\newcommand{\tR}{ \textsc{r} }

\begin{document}

\title{Chiral matching of dimension-7 baryon-number-violating operators and application to $\Delta I=3/2$ nucleon decays}
\author{Xiao-Dong Ma\,\orcidlink{0000-0001-7207-7793}}
\email{maxid@scnu.edu.cn}
\affiliation{State Key Laboratory of Nuclear Physics and
Technology, Institute of Quantum Matter, South China Normal
University, Guangzhou 510006, China}
\affiliation{Guangdong Basic Research Center of Excellence for
Structure and Fundamental Interactions of Matter, Guangdong
Provincial Key Laboratory of Nuclear Science, Guangzhou
510006, China}

\begin{abstract}

We study chiral matching of dimension-7 (dim-7) baryon-number-violating operators in low-energy effective field theory. 
We show that all dim-7 operators can be classified into two distinct Lorentz structures and their chirality-flipped counterparts, and we derive their leading-order hadronic realizations in chiral perturbation theory. As an application, we systematically investigate $\Delta I=3/2$ nucleon decays, including $n\to \ell^-\pi^+$,  
$p\to \ell^- \pi^+\pi^+$, and  
$n\to \ell^- \pi^+ (\pi^0, \eta)$,
with $\ell=e,~\mu$. 
Using existing experimental bounds on $n\to \ell^-\pi^+$, we set stringent constraints on the relevant Wilson coefficients and derive improved lower limits on the partial lifetimes of the six three-body modes from their correlations with the two-body modes. 
We further present a concrete ultraviolet model that generates the dim-7 operators and the associated $\Delta I=3/2$ nucleon decays at leading order.
Our results pave the way for exploring dim-7 contributions to nucleon decays and motivate future searches for these exotic modes.
\end{abstract}

\maketitle 

\section{Introduction}

Nucleon decay searches are among the primary physics goals of ongoing and next-generation large-fiducial-mass neutrino experiments, including Super-Kamiokande (Super-K), JUNO~\cite{JUNO:2015zny}, Hyper-Kamiokande~\cite{Hyper-Kamiokande:2018ofw}, DUNE~\cite{DUNE:2020ypp}, and Theia~\cite{Theia:2019non}.
Currently, Super-K is the leading experiment in the search for various nucleon decay modes~\cite{Super-Kamiokande:2025lxa,Super-Kamiokande:2026yep,Kamiokande:2026net}, continuously improving its sensitivity and setting increasingly stringent limits on their occurrence. JUNO began data taking in 2025~\cite{JUNO:2025gmd,JUNO:2025fpc} and is expected to release its first nucleon decay search results in the near future. 

Nucleon decay is a low-energy process occurring below the GeV scale, where a general and model-independent description can be formulated within the framework of effective field theory (EFT). In particular, the combined framework of low-energy effective field theory (LEFT)~\cite{Jenkins:2017jig,Liao:2020zyx} and baryon chiral perturbation theory (ChPT)~\cite{Jenkins:1990jv,Bijnens:1985kj,Claudson:1981gh,Liao:2025vlj,Liao:2025sqt}
provides a systematic treatment of a wide range of nucleon decay processes.
Within this framework, both conventional and exotic channels involving only standard model (SM) particles have been extensively explored in recent works~\cite{Beneito:2023xbk,Gargalionis:2024nij,Chen:2025mjt,Liao:2025wxk,Fan:2026fqo,Fan:2026csl}. 
These studies have mainly focused on non-derivative baryon-number-violating (BNV) operators, including dimension-6 (dim-6) four-fermion and dim-9 six-fermion operators. 

Although nucleon decays induced by dim-6 LEFT operators have been extensively studied, 
comparatively less attention has been devoted to 
dim-7 BNV operators containing an additional derivative, even though they can generate certain decay channels at leading order. Their contributions are particularly relevant when those from dim-6 operators are suppressed by symmetries or cancellations in the underlying theory.
Notably, decay modes involving an isospin change of $\Delta I=3/2$ can arise at leading order only from dim-7 or higher-dimensional operators, including the two- and three-body decays $n\to \ell^-\pi^+$, $p\to \ell^-\pi^+\pi^+$, and $n\to \ell^-\pi^+(\pi^0,\eta)$, with $\ell=e,\,\mu$. 
This follows from the fact that dim-6 operators can induce transitions with an isospin change of at most one unit~\cite{Liao:2025vlj}. 
 
In this work, we aim to systematically match the dim-7 BNV LEFT operators onto ChPT and study the exotic $\Delta I=3/2$ nucleon decays they induce. We find that all dim-7 BNV operators can be organized into two distinct Lorentz structures and their chirality-flipped partners, each with a unique leading-order chiral realization in ChPT. We then apply this formalism to derive decay widths of the $\Delta I=3/2$ modes in terms of the Wilson coefficients (WCs) of the relevant dim-7 LEFT and SMEFT operators~\cite{Liao:2016hru,Lehman:2014jma}. 
Using existing IMB limits on the two-body modes $n\to (e^-,\mu^-)\pi^+$~\cite{Seidel:1988ut}, we obtain stringent constraints on the relevant WCs and, consequently, improved indirect limits on the three-body modes. 

The remainder of this paper is organized as follows. 
\cref{sec:dim7LEFT} summarizes the dim-7 BNV operators in LEFT, followed by their chiral realizations in \cref{sec:chiralmatching}. 
As a concrete application of the chiral matching results, \cref{sec:n2lmpip} presents a systematic investigation of the $\Delta I=3/2$ nucleon decays $n\to \ell^-\pi^+$, $p\to \ell^-\pi^+\pi^+$, and $n\to \ell^-\pi^+(\pi^0,\eta)$.
A renormalizable ultraviolet (UV) model that generates these dim-7 derivative-type operators and the corresponding $\Delta I=3/2$ nucleon decays at leading order is provided in \cref{sec:uvmodels}. 
We conclude in \cref{sec:summary}.
Additional details on operator-basis conversion and expressions for spurion fields are collected in Appendices \ref{app:bases_trans} and \ref{app:spurions}, respectively. 

\section{Dimension-7 BNV LEFT operators}
\label{sec:dim7LEFT}

The LEFT is a tailored effective field theory for low-energy processes below the electroweak scale, making it particularly well suited for the systematic study of BNV processes such as nucleon decays.
Its field content consists of the five light quarks $(u,d,s,c,b)$, three charged leptons and their associated neutrinos $(e,\mu,\tau,\nu_e,\nu_\mu,\nu_\tau)$, and the gauge bosons (the photon and gluons).
The higher-dimensional operators constructed from these fields are required to be invariant under the QCD and QED gauge symmetries $\rm SU(3)_c\otimes U(1)_{\rm em}$. The leading BNV operators appear at dimension 6, with 9 and 7 flavor-blind non-Hermitian operators (i.e., without accounting for the lepton and quark flavors) in the $\Delta(B-L)=0$ and $\Delta(B+L)=0$ sectors, respectively~\cite{Jenkins:2017jig}.

At dimension 7, the relevant BNV operators contain an additional derivative acting on the fermion fields. 
The complete dim-7 BNV operator basis is given in~\cite{Liao:2020zyx}, comprising 9 and 7 flavor-blind operators in the $\Delta(B-L)=0$ and  $\Delta(B+L)=0$ sectors, respectively. 
To facilitate the later chiral matching, we employ a modified basis in which quark fields of the same chirality are fully symmetrized with respect to their flavor indices. 
For convenience, we collectively denote the three charged leptons by $\ell=e,\mu,\tau$, three left-handed neutrinos by $\nu_\tL=\nu_{e\tL},\nu_{\mu\tL},\nu_{\tau\tL}$, the three down-type quarks by $\text{d}=d,s,b$, and the two light up-type quarks by $\text{u}=u,c$.
The explicit parametrization of the 16 operators used in this work is given below:
\begin{subequations}
\label{eq:dim7basis}
\begin{align}
\calO_{\partial\ell\rm uud}^{\tL\tR} & =
i(\overline{D_\mu\ell_\tL^\C}\{\text{u}_\tL^\alpha)
(\overline{\text{u}_\tL^{\beta\C}}\}\gamma^\mu
\text{d}_\tR^\gamma) \epsilon_{\alpha\beta\gamma}
\sim \calO_{deuD1}, 
\nonumber\\
\calO_{\partial\ell\rm udu}^{\tL\tR} & =
i(\overline{D_\mu\ell_\tL^\C}\{\text{u}_\tL^\alpha)
(\overline{\text{d}_\tL^{\beta\C}}\}
\gamma^\mu \text{u}_\tR^\gamma) \epsilon_{\alpha\beta\gamma}
\sim \calO_{ueudD1}, 
\nonumber\\
\calO_{\partial\nu\rm ddu}^{\tL\tR} & =
i(\overline{\partial_\mu\nu_\tL^\C}\{\text{d}_\tL^\alpha)
(\overline{\text{d}_\tL^{\beta\C}}\}\gamma^\mu
\text{u}_\tR^\gamma) \epsilon_{\alpha\beta\gamma}
\sim \calO_{u\nu dD3}, 
\nonumber\\
\calO_{\partial\nu\rm udd}^{\tL\tR} & =
i(\overline{\partial_\mu\nu_\tL^\C}\{\text{u}_\tL^\alpha)
(\overline{\text{d}_\tL^{\beta\C}}\} \gamma^\mu
\text{d}_\tR^\gamma) \epsilon_{\alpha\beta\gamma}
\sim \calO_{d\nu udD2}, 
\nonumber\\
\calO_{\partial\bar\ell\rm ddd}^{\tL\tR} & =
i(\overline{D_\mu\ell_\tR}\{\text{d}_\tL^\alpha)
(\overline{\text{d}_\tL^{\beta\C}}\}\gamma^\mu
\text{d}_\tR^\gamma) \epsilon_{\alpha\beta\gamma}
\sim \calO_{dedD3}^\dagger,
\\%
\calO_{\partial\ell\rm uud}^{\tR\tL} & =
i(\overline{D_\mu\ell_\tR^\C}\{\text{u}_\tR^\alpha)
(\overline{\text{u}_\tR^{\beta\C}}\}\gamma^\mu
\text{d}_\tL^\gamma) \epsilon_{\alpha\beta\gamma}
\sim \calO_{deuD4},
\nonumber\\
\calO_{\partial\ell\rm udu}^{\tR\tL} & =
i(\overline{D_\mu\ell_\tR^\C}\{\text{u}_\tR^\alpha)
(\overline{\text{d}_\tR^{\beta\C}}\}
\gamma^\mu \text{u}_\tL^\gamma) \epsilon_{\alpha\beta\gamma}
\sim \calO_{ueudD2}, 
\nonumber\\
\calO_{\partial\bar\nu\rm ddu}^{\tR\tL} & =
i(\overline{\partial_\mu\nu_\tL}\{\text{d}_\tR^\alpha)
(\overline{\text{d}_\tR^{\beta\C}}\}\gamma^\mu
\text{u}_\tL^\gamma) \epsilon_{\alpha\beta\gamma}
\sim \calO_{u\nu dD2}^\dagger,
\nonumber\\
\calO_{\partial\bar\nu\rm udd}^{\tR\tL} & =
i(\overline{\partial_\mu\nu_\tL}\{\text{u}_\tR^\alpha)
(\overline{\text{d}_\tR^{\beta\C}}\} \gamma^\mu
\text{d}_\tL^\gamma) \epsilon_{\alpha\beta\gamma}
\sim \calO_{d\nu udD1}^\dagger,
\nonumber\\
\calO_{\partial\bar\ell\rm ddd}^{\tR\tL} & =
i(\overline{D_\mu\ell_\tL}\{\text{d}_\tR^\alpha)
(\overline{\text{d}_\tR^{\beta\C}}\}\gamma^\mu
\text{d}_\tL^\gamma) \epsilon_{\alpha\beta\gamma}
\sim \calO_{dedD2}^\dagger,
\\%
\calO_{\partial\ell\rm duu}^{\tL\tL} & = 
(\overline{\ell_\tR^\C}\gamma^\mu\{\text{d}_\tL^\alpha)
(\overline{\text{u}_\tL^{\beta\C}}i\overleftrightarrow{D_\mu}
\text{u}_\tL^\gamma\}) \epsilon_{\alpha\beta\gamma}
\sim \calO_{deuD3},
\nonumber\\
\calO_{\partial\bar\nu\rm udd}^{\tL\tL} & =
(\overline{\nu_\tL}\gamma^\mu \{\text{u}_\tL^\alpha)
(\overline{\text{d}_\tL^{\beta\C}}i\overleftrightarrow{D_\mu}
\text{d}_\tL^\gamma\}) \epsilon_{\alpha\beta\gamma} 
\sim \calO_{u\nu dD1}^\dagger, 
\nonumber\\
\calO_{\partial\bar\ell\rm ddd}^{\tL\tL} & =
(\overline{\ell_\tL}\gamma^\mu \{\text{d}_\tL^\alpha)
(\overline{\text{d}_\tL^{\beta\C}}i \overleftrightarrow{D_\mu}
\text{d}_\tL^\gamma\}) \epsilon_{\alpha\beta\gamma}  
\sim \calO_{dedD1}^\dagger, 
\\%
\calO_{\partial\ell\rm duu}^{\tR\tR} & =
(\overline{\ell_\tL^\C}\gamma^\mu \{\text{d}_\tR^\alpha)
(\overline{\text{u}_\tR^{\beta\C}}i\overleftrightarrow{D_\mu}
\text{u}_\tR^\gamma\}) \epsilon_{\alpha\beta\gamma}
\sim \calO_{deuD2},
\nonumber\\
\calO_{\partial\nu\rm udd}^{\tR\tR} & =
(\overline{\nu_\tL^\C}\gamma^\mu\{\text{u}_\tR^\alpha)
(\overline{\text{d}_\tR^{\beta\C}}i \overleftrightarrow{D_\mu}
\text{d}_\tR^\gamma\}) \epsilon_{\alpha\beta\gamma}   
\sim \calO_{u\nu dD4},
\nonumber\\
\calO_{\partial\bar\ell\rm ddd}^{\tR\tR} & =
(\overline{\ell_\tR}\gamma^\mu\{\text{d}_\tR^\alpha)
(\overline{\text{d}_\tR^{\beta\C}}i \overleftrightarrow{D_\mu}
\text{d}_\tR^\gamma\}) \epsilon_{\alpha\beta\gamma}
\sim \calO_{dedD4}^\dagger,
\end{align}
\end{subequations}
where $A\overleftrightarrow{D_\mu}B=A (D_\mu B) - (D_\mu A)B$.
The left- and right-handed chiral fields are denoted by the subscripts $\tL$ and $\tR$, respectively, while charge conjugation of a fermion field $\psi$ is indicated by $\psi^\C$.
The curly brackets indicate total symmetrization over the quark flavors inside, which are left implicit. 
When an operator exhibits flavor symmetries, we retain only one flavor-specific representative for each independent case at the Lagrangian level. 
For example, for the operator $\calO_{\partial\ell\rm duu}^{\tL\tL}$, the flavor assignments $\text{duu}=duc$ and $\text{duu}=dcu$ are equivalent. 
We therefore retain the flavor-specific operator $\calO_{\partial\ell duc}^{\tL\tL}$ whose specific form is 
$\calO_{\partial\ell duc}^{\tL\tL} =
(1/6)[ (\overline{\ell_\tR^\C}\gamma^\mu d_\tL^\alpha)
(\overline{u_\tL^{\beta\C}}i\overleftrightarrow{D_\mu}
c_\tL^\gamma) \epsilon_{\alpha\beta\gamma}  
+\mbox{5 perms of~}(d,u,c)]$.

In the rightmost column of \cref{eq:dim7basis}, 
we indicate the correspondence with the operator conventions adopted in~\cite{Liao:2020zyx}.
The two operator bases can be transformed into each other by using Fierz identities, integration by parts, and the equations of motion of fermion fields. 
An explicit example demonstrating this equivalence is provided in Appendix \ref{app:bases_trans}.
The operators in the $\Delta(B-L)=0$ and 
$\Delta(B+L)=0$ sectors are distinguished by the absence or presence of a bar over the lepton field in the operator notation, respectively. 

To study the leading contributions to nucleon decays, we restrict our attention to operators involving only the light $u,d,s$ quarks. 
Without counting lepton flavors, 
\cref{eq:dim7basis} contains 17\,+\,17 operators with $\tL\tR+\tR\tL$ structures and 9\,+\,9 operators with $\tL\tL+\tR\tR$ structures. 
For convenience, we define the triple-quark operators corresponding to the four distinct Lorentz structures as follows:
\begin{subequations}
\label{eq:Nyzw}
\begin{align}
\calN_{yzw}^{\tL\tR,\mu} & = q_{\tL,\{y}^\alpha (\overline{ q_{\tL, z\}}^{\beta \C} } \gamma^\mu q_{\tR,w}^\gamma)\epsilon_{\alpha \beta \gamma},
\\
\calN_{yzw}^{\tL\tL,\mu} & =  q_{\tL, \{y}^\alpha (\overline{ q_{\tL, z}^{\beta \C} } i \overleftrightarrow{D^\mu} q_{\tL, w\} }^\gamma )\epsilon_{\alpha \beta \gamma},
\end{align}
\end{subequations} 
plus their chirality-flipped partners with $\tL\leftrightarrow\tR$. 
Here, $y,z,w=1,2,3$ are light quark flavor indices, with $q_{1,2,3}=u,d,s$. 
The symmetrizations over quark flavors are defined as 
$A_{\{y}B_{z\}} = (1/2)(A_{y} B_{z} + A_{z}B_{y})$ and
$A_{\{y}B_z C_{w\}} = (1/6) [ A_{y}B_{z} C_w + \mbox{5 perms of~}(y,z,w)]$.
The structures $\calN_{yzw}^{\tL\tR(\tR\tL),\mu}$ were recently identified in~\cite{Liao:2025vlj}.
Then all 52 operators can be compactly organized as follows:
\begin{subequations}
\label{eq:Ope_dlqqq}
\begin{align}
{\cal O}_{\partial l yzw}^{\tL\tR} & = i\overline{D_\mu l}\calN_{yzw}^{\tL\tR,\mu},
& 
{\cal O}_{\partial l yzw }^{\tR\tL} & = i\overline{D_\mu l}\calN_{yzw}^{\tR\tL,\mu},
\\
{\cal O}_{\partial l yzw}^{\tL\tL} & = \overline{l}\gamma_\mu \calN_{yzw}^{\tL\tL,\mu}, 
&
{\cal O}_{\partial l yzw}^{\tR\tR} & = \overline{l}\gamma_\mu \calN_{yzw}^{\tR\tR,\mu}, 
\end{align}
\end{subequations}
where the lepton field 
$l = \ell_{\tL(\tR)}, 
\ell_{\tL(\tR)}^\C, \nu_{\tL}, \nu_\tL^\C$ depending on the quark configuration.
For each operator $\calO_i^j$, we define a spurion field as the product of its non-QCD component and the associated WC $C_i^j$ by $\calP_i^j$. 
The general dim-7 BNV LEFT Lagrangian involving $u,d,s$ quarks then takes the form:
\begin{align}
{\cal L}_{\partial l q^3}^{\slashed{B}} & =
 \big[
\calP_{yzw}^{\tL\tR,\mu}
\calN_{yzw,\mu}^{\tL\tR}
+ \calP_{yzw}^{\tR\tL,\mu}
\calN_{yzw,\mu}^{\tR\tL}
\big]
\nonumber 
\\%
& + \big[
\calP_{yzw}^{\tL\tL,\mu}
\calN_{yzw,\mu}^{\tL\tL}
+ 
\calP_{yzw}^{\tR\tR,\mu}
\calN_{yzw,\mu}^{\tR\tR}
\big]+\text{H.c.},
\label{eq:Lag_lqqqD}
\end{align}
where summation over quark flavor indices $y,z,w$ and lepton flavor is implied. 
The spurion fields associated with the 52 quark-flavor-specific operators are summarized in Appendix \ref{app:spurions}. 

The effective operators ${\cal O}_i^{yzw}$ in \cref{eq:Ope_dlqqq}, together with the Lagrangian in \cref{eq:Lag_lqqqD}, are defined at the 
chiral symmetry breaking scale $\Lambda_{\chi}\approx 1.2~\rm GeV$, below which the nonperturbative QCD effects become significant.
However, these LEFT operators are typically matched at the electroweak (EW) scale $\Lambda_{\rm EW}\approx160~\rm GeV$, to the corresponding operators in the SMEFT or its extensions. Consequently, their contributions to nucleon decay are subject to renormalization group (RG) evolution between $\Lambda_{\rm EW}$ and $\Lambda_{\chi}$, with the leading corrections arising from one-loop QCD effects. We find that the one-loop RG equations are 
\begin{align}
\frac{d C_{\partial l yzw}^{\tL\tR(\tR\tL)} }{d\ln\mu} =  \frac{2}{3}\frac{\alpha_s}{2\pi} C_{\partial l yzw}^{\tL\tR(\tR\tL)},\quad 
\frac{d C_{\partial l yzw}^{\tL\tL(\tR\tR)} }{d\ln\mu} =  2\frac{\alpha_s}{2\pi} C_{\partial l yzw}^{\tL\tL(\tR\tR)},
\end{align}
where $\alpha_s$ denotes the strong coupling constant.
Taking into account the one-loop RG running of $\alpha_s$, we obtain numerically,
$C_{\partial l yzw}^{\tL\tR(\tR\tL)}(\Lambda_\chi)\approx 
0.91\,C_{\partial l yzw}^{\tL\tR(\tR\tL)}(\Lambda_{\rm EW})$ 
and 
$ C_{\partial l yzw}^{\tL\tL(\tR\tR)}(\Lambda_\chi)\approx 
0.76\,C_{\partial l yzw}^{\tL\tL(\tR\tR)}(\Lambda_{\rm EW})$.

We are now in a position to analyze the transformation properties of the triple-quark structures in \cref{eq:Nyzw} and their chirality-flipped partners under the chiral symmetry group $G_\chi=\rm SU(3)_\tL\otimes SU(3)_\tR$ of the light $u,d,s$ quarks.
Inspecting the chiral quark fields, one can readily obtain their irreducible representations under $G_\chi$:
\begin{subequations}
\begin{align}
\calN_{yzw}^{\tL\tR,\mu} & \in \pmb{6}_\tL \otimes \pmb{3}_\tR,
& \calN_{yzw}^{\tR\tL,\mu} & \in \pmb{3}_\tL \otimes \pmb{6}_\tR,
\\
\calN_{yzw}^{\tL\tL,\mu} &\in \pmb{10}_\tL \otimes  \pmb{1}_\tR, 
& \calN_{yzw}^{\tR\tR,\mu} &\in \pmb{1}_\tL \otimes  \pmb{10}_\tR. 
\end{align}
\end{subequations} 
For the subsequent chiral matching, 
it is convenient to promote the effective Lagrangian in \cref{eq:Lag_lqqqD} to a formally chiral-invariant form by assigning appropriate chiral transformation properties to the spurion fields. 
Under the chiral transformations
$q_{\tL} \to \hat L  q_\tL$ and 
$q_{\tR} \to \hat R  q_\tR$, where 
$(\hat L,\hat R)\in G_\chi$, the corresponding transformation rules for the spurion fields are 
\begin{subequations}
\label{eq:spurion_trans}
\begin{align} 
{\cal P}_{yzw}^{\tL\tR,\mu}
& \to
\hat L^*_{yy'} \hat L^*_{zz'} \hat R_{ww'}^*  
{\cal P}_{y'z'w'}^{\tL\tR,\mu},
\\%
{\cal P}_{yzw}^{\tL\tL,\mu}
& \to \hat L_{yy'}^* \hat L_{zz'}^* \hat L_{ww'}^*
{\cal P}_{y'z'w'}^{\tL\tL,\mu}.
\end{align} 
\end{subequations} 
The transformation rules for the corresponding chirality-flipped partners are obtained by simultaneously interchanging 
$\tL\leftrightarrow\tR$ and 
$\hat L \leftrightarrow\hat R$.

\section{Chiral matching}
\label{sec:chiralmatching}

To calculate the nucleon decay matrix elements induced by the effective interactions in \cref{eq:Lag_lqqqD}, 
we employ ChPT to match the triple-quark operators onto their corresponding hadronic representations, composed of the baryon and pseudoscalar meson fields. 
In addition to the spurion fields $\calP_i^j$, 
we arrange the pseudoscalar meson and baryon octet fields into the following matrix form:
\begin{subequations}
\begin{align}
\Sigma(x) & = \xi^2(x) = \exp\Big[\frac{i\sqrt{2}\Pi(x)}{F_0}\Big],  
\\
\Pi(x) & =   
\begin{pmatrix}
\frac{\pi^0}{\sqrt{2}}+\frac{\eta}{\sqrt{6}} & \pi^+ & K^+\\
\pi^- & -\frac{\pi^0}{\sqrt{2}}+\frac{\eta}{\sqrt{6}} & K^0\\
K^- & \bar{K}^0 & -\sqrt{\frac{2}{3}}\eta
\end{pmatrix},
\\
B(x) &=
\begin{pmatrix}
\frac{\Sigma^{0}}{\sqrt{2}}+\frac{\Lambda^0}{\sqrt{6}}  & \Sigma^+ & p \\
\Sigma^- & -\frac{\Sigma^{0}}{\sqrt{2}}
+\frac{\Lambda^0}{ \sqrt{6}} &  n \\ 
\Xi^- & \Xi^0 & - \sqrt{\frac{2}{3}}\Lambda^0
\end{pmatrix},   
\end{align}
\end{subequations}
where $F_0$ denotes the pion decay constant in the chiral limit. 
Numerically, $F_0 = f_{\pi}/\sqrt{2}$ with $f_\pi=130.41(20)\,\rm MeV$~\cite{ParticleDataGroup:2024cfk}.
Under the chiral transformations $(\hat L,\hat R)\in G_\chi$,
these fields transform as 
$\Sigma \to \hat L \Sigma \hat R^\dagger$, 
$ B \to \hat h B \hat h^\dagger$, 
$\xi \to \hat L \xi \hat h^\dagger = \hat h \xi \hat R^\dagger$, 
where $\hat h$ is a compensating transformation that depends on $\hat L$, $\hat R$, and $\xi$. 
These transformation properties imply that   
$\xi B \xi \to 
\hat L(\xi B \xi) \hat R^\dagger$,  
$\xi^\dagger B \xi^\dagger \to 
\hat R (\xi^\dagger B \xi^\dagger) \hat L^\dagger$, 
$\xi B \xi^\dagger \to 
\hat L (\xi B \xi^\dagger) \hat L^\dagger$,
and $\xi^\dagger B \xi \to 
\hat R (\xi^\dagger B \xi) \hat R^\dagger$, respectively. 

Together with the transformation rules for the spurion fields given in \cref{eq:spurion_trans}, 
the corresponding chiral-invariant Lagrangian can be constructed from the spurion fields, the meson and baryon octet fields, and their covariant derivatives. 
To identify the leading-order chiral operators, 
we adopt the chiral derivative power-counting scheme in which  
$\{\Sigma, \xi, B,D_\mu B\}\sim {\cal O}(p^0)$ and 
$D_\mu\Sigma \sim {\cal O}(p^1)$.
The derivatives $D_\mu \Sigma$ and $D_\mu B$ 
are defined as $D_\mu \Sigma = \partial_\mu \Sigma -il_{\mu}\Sigma+i\Sigma r_{\mu}$ and $D_\mu B = \partial_\mu B + [\Gamma_\mu ,B]$, where $\Gamma_{\mu}=\frac{1}{2}\left[\xi(\partial_{\mu}-ir_{\mu})\xi^{\dagger}+\xi^{\dagger}(\partial_{\mu}-il_{\mu})\xi\right]$.
Here, the external sources $l_{\mu}$ and $r_{\mu}$, which couple to the quark currents $\overline{q_\tL}\gamma^\mu q_\tL$ and $\overline{q_\tR}\gamma^\mu q_\tR$, respectively, are traceless matrices in flavor space.

The chiral matching for the irreps $\pmb{6}_{\tL(\tR)} \otimes \pmb{3}_{\tR(\tL)}$ was worked out in~\cite{Liao:2025vlj} and 
can be written as: 
\begin{align}
{\cal L}_{\pmb{6}\otimes\pmb{3}}^{\slashed{B}} =
& \frac{c_3}{\Lambda_\chi} \big[ 
\calP_{yzi}^{\tL\tR,\mu}
{\Gamma}_{\mu\nu}^{\tt L} 
(\xi i D^\nu B_\tL \xi)_{yj}
\Sigma_{zk} \epsilon_{ijk}
\nonumber
\\
&- \calP_{yzi}^{\tR\tL,\mu}
{\Gamma}_{\mu\nu}^{\tt R}
(\xi^\dagger i D^\nu B_\tR  \xi^\dagger)_{yj}
\Sigma^*_{kz} \epsilon_{ijk} \big]
+\text{H.c.},
\label{eq:L6x3}
\end{align}%
where $\Gamma_{\mu\nu}^{\tL(\tR)} = (g_{\mu\nu} - {1\over 4} \gamma_\mu \gamma_\nu)P_{\tL(\tR)}$ are Lorentz projectors introduced to preserve the constraints $\gamma_\mu \calN_{yzw}^{\tL\tR(\tR\tL),\mu}=0$ at the hadronic level.
For the totally symmetric irreps 
$\pmb{10}_{\tL(\tR)}\otimes \pmb{1}_{\tR(\tL)}$, the leading chiral matching is likewise unique.
Unlike the $\pmb{6}_{\tL(\tR)}\otimes \pmb{3}_{\tR(\tL)}$ case, however, it necessarily involves a derivative acting on the octet meson field. 
Consequently, this matching takes the form: 
\begin{align}
{\cal L}_{\pmb{10}\otimes\pmb{1}}^{\slashed{B}} & =  
c_5  \big[ 
\calP_{yzw}^{\tL\tL,\mu}
(\xi B_\tL \xi)_{yi}
\Sigma_{zj}i(D_{\mu}\Sigma)_{wk} \epsilon_{ijk}
\nonumber
\\
& - \calP_{yzw}^{\tR\tR,\mu} 
(\xi^\dagger B_\tR \xi^\dagger)_{yi}
\Sigma^*_{jz} i(D_{\mu}\Sigma)^*_{kw} \epsilon_{ijk}\big]
+\text{H.c.}.
\label{eq:L10x1}
\end{align}
The minus sign in \cref{eq:L6x3,eq:L10x1} is required 
to ensure the correct parity relation between the triple-quark operators and their corresponding hadronic counterparts. 

The coefficients $c_{3,5}$ are low-energy constants (LECs) that have not yet been determined from lattice QCD or experiment. 
A naive dimensional analysis (NDA) estimation gives $c_3 \approx 0.011\,{\rm GeV}^3$~\cite{Liao:2025vlj}. 
We estimate $c_{5}$ similarly,
following Weinberg's approach~\cite{Weinberg:1989dx}
by equating the reduced couplings $C_q$ and $C_h$ of the quark- and hadron-level operators.
For an interaction term with coupling $g$ and a dim-$D$ operator containing $m$ physical fields, the reduced coupling is defined as $g(4\pi)^{2-m} \Lambda_\chi^{D-4}$. 
Omitting the spurion fields, 
the derivative-type triple-quark operator $\calN_{yzw}^{\tL\tL,\mu}$ has $g=1$, $m=3$, and $D=11/2$, yielding 
$C_q = (4\pi)^{-1} \Lambda_\chi^{3/2}$. 
Expanding the corresponding hadronic operator in \cref{eq:L10x1} to its first nonvanishing term,  $(\sqrt{2}c_5/F_0)B \partial M$, gives $g=\sqrt{2}c_5/F_0$, $m=2$, and $D=7/2$,
and hence $C_h = (\sqrt{2} c_5/F_0) \Lambda_\chi^{-1/2}$.  
Requiring $C_h=C_q$ then gives   
$c_5 \sim \Lambda_\chi^2 F_0 /(4\sqrt{2}\pi) \approx 0.007\,{\rm GeV}^3$. 

We now make a few comments on the results in \cref{eq:L6x3,eq:L10x1}. 
First, the above matching results can be extended to BNV operators beyond dimension 7: for operators sharing the same triple-quark structures as those in \cref{eq:Nyzw}, their leading-order chiral realizations are obtained by appropriately replacing the spurion fields in \cref{eq:L6x3,eq:L10x1}.
Second, we neglect contributions involving baryon decuplet or vector mesons, which are typically subleading as demonstrated in~\cite{Liao:2025vlj,Fan:2026fqo,Fan:2026csl}.
We also neglect higher-order chiral corrections, including operators with additional derivatives or quark-mass insertions, which are expected to be small. 
Third, the chiral matching of the dim-7 operators was also studied recently in~\cite{Song:2026gyo}, using a different operator basis and chiral construction method, making a direct comparison with our results difficult. 
We note, however, that their leading-order realization of the operators in the irreps $\pmb{6}_{\tL(\tR)}\times \pmb{3}_{\tR(\tL)}$ occurs at higher chiral order and therefore does not generate the baryon-lepton pole terms present in our analysis. From the perspective of flavor symmetry, however, it is not apparent why such direct baryon-pole contributions should be forbidden. 
Finally, substituting the spurion fields in Appendix \ref{app:spurions} into \cref{eq:L6x3,eq:L10x1} and expanding the pseudoscalar meson fields to the required order yields the local vertices $Bl M^n$ ($n=0,1,2,\cdots$). 
Combined with the standard baryon ChPT Lagrangian~\cite{Jenkins:1990jv,Bijnens:1985kj}, these vertices provide the ingredients needed to calculate the relevant nucleon decay amplitudes. 
In the following section, we use this framework to systematically analyze $\Delta I=3/2$ nucleon decays.

\section{$\Delta I=3/2$ nucleon decays}
\label{sec:n2lmpip}

The two- and three-body decay modes $n\to \ell^- \pi^+$, $p\to \ell^- \pi^+\pi^+$, and $n\to \ell^- \pi^+(\pi^0,\eta)$, with $\ell=e,\mu$, require a change in the isospin of $\Delta I=3/2$. 
At leading order, these modes are mediated by four dim-7 BNV operators involving three down quarks:   
\begin{align}
\calO_{\partial\bar\ell ddd}^{\tR\tL} &= i (\overline{D_\mu\ell_\tL} d_\tR^\alpha)(\overline{d_\tR^{\beta\C} }\gamma^\mu d_\tL^\gamma)
\epsilon_{\alpha\beta\gamma},  
\nonumber \\
\calO_{\partial\bar\ell ddd}^{\tR\tR} &= (\overline{\ell_\tR} \gamma_\mu d_\tR^\alpha)(\overline{d_\tR^{\beta\C} }i \overleftrightarrow{D^\mu} d_\tR^\gamma)
\epsilon_{\alpha\beta\gamma}, 
\end{align}
together with their chirality-flipped partners with $\tL\leftrightarrow \tR$. 
The two-pion modes can also receive contributions from $\Delta I=5/2$ dim-9 operators, which we neglect due to their higher dimensionality.
From Appendix \ref{app:spurions}, the corresponding nonvanishing spurion fields are
\begin{align}
\calP_{ddd}^{\tR\tL,\mu} =\,& i C_{\partial\bar\ell ddd}^{\tR\tL} \overline{D^\mu\ell_\tL}, &
\calP_{ddd}^{\tL\tR,\mu} =\,& i C_{\partial\bar\ell ddd}^{\tL\tR} \overline{D^\mu\ell_\tR}, 
\nonumber \\
\calP_{ddd}^{\tR\tR,\mu} =\,& C_{\partial\bar\ell ddd}^{\tR\tR} \overline{\ell_\tR}\gamma^\mu, &
\calP_{ddd}^{\tL\tL,\mu} =\,& C_{\partial\bar\ell ddd}^{\tL\tL} \overline{\ell_\tL}\gamma^\mu. 
\end{align}
By expanding the chiral Lagrangian terms in \cref{eq:L6x3,eq:L10x1} to linear order in the meson fields and taking the spurion fields into consideration, 
we obtain the relevant three-point interactions involving a neutron, a pion, and a charged lepton, 
\begin{align}
{\cal L}_{\bar\ell n \pi} =&\, 
 - i \frac{\sqrt{2}c_{3\chi} }{F_0} C_{\partial\bar\ell ddd}^{\tR\tL}\,
 \pi^- (\overline{\partial^\mu\ell_\tL}\tilde{\partial}_\mu n_\tR)
\nonumber\\
&- i  \frac{\sqrt{2}c_5}{F_0 }   C_{\partial\bar\ell ddd}^{\tR\tR}\,
 i \partial_\mu  \pi^- (\overline{\ell_\tR}\gamma^\mu n_\tR) + \tL \leftrightarrow \tR,
 \label{eq:Lnlpi}
\end{align}
where $c_{3\chi}=c_3/\Lambda_\chi$ and $\tilde{\partial}_\mu=\partial_\mu-\frac{1}{4}\gamma_\mu\slashed{\partial}$.
By expanding the chiral Lagrangian terms to the second order in the meson fields, we obtain the following four-point vertices related to the three-body modes, 
\begin{align}
{\cal L}_{\bar\ell\tN \pi M} =&\,  
- \frac{\sqrt{6}c_{3\chi}}{6F_0^2}  C_{\partial\bar\ell ddd}^{\tR\tL} (\pi^- \eta - \sqrt{3} \pi^- \pi^0)
 (\overline{\partial^\mu\ell_\tL} \tilde{\partial}_\mu n_\tR) 
\nonumber\\
& - \frac{ c_{3\chi}}{F_0^2} C_{\partial \bar\ell ddd}^{\tR\tL} \pi^-\pi^- ( \overline{\partial^\mu\ell_\tL}\tilde{\partial}_\mu p_\tR)
\nonumber\\
&+\frac{\sqrt{2}  c_{5}}{2 F_0^2} C_{\partial\bar\ell ddd}^{\tR\tR} (3 \pi^0 i\partial_\mu\pi^-   - 2 \pi^- i\partial_\mu\pi^0 
\nonumber\\
&- \sqrt{3} \eta i\partial_\mu\pi^-) (\overline{\ell_\tR}\gamma^\mu n_\tR) 
\nonumber\\
&
- \frac{c_{5}}{F_0^2} C_{\partial\bar\ell ddd}^{\tR\tR}  \pi^- i \partial_\mu \pi^- (\overline{\ell_\tR}\gamma^\mu p_\tR)
- \tL\leftrightarrow \tR.
\end{align}

In addition to the BNV interactions given above, the standard leading-order chiral interactions involving baryons are also required for the three-body modes through noncontact diagrams. These interactions have been worked out previously and we summarize the relevant terms as follows~\cite{Fan:2026csl},
\begin{align}
\label{eq:LBBM}
{\cal L}_{\bar BBM} \supset &  
\frac{D+F}{\sqrt{2} F_0} (\overline{n} \gamma^\mu \gamma_5 p) \partial_\mu \pi^-
-\frac{D+F}{2 F_0} (\overline{n} \gamma^\mu \gamma_5 n) \partial_\mu \pi^0
\nonumber\\
&+\frac{3F-D}{2\sqrt{3} F_0} (\overline{n} \gamma^\mu \gamma_5 n) \partial_\mu \eta.
\end{align}
Numerically, we use the low-energy constants $D=0.730(11)$ and $F\approx0.447$ from the lattice calculation~\cite{Bali:2022qja}.

\subsection{Two-body decays $n\to \ell^- \pi^+$ }

\begin{figure}[ht]
\centering
\begin{tikzpicture}[mystyle,scale=1]
\begin{scope}
\draw[f] (0, 0)node[left]{$n$} -- (1.5,0);
\draw[f] (1.5, 0) -- (3,0) node[right]{$\ell^-$};
\draw[snar, black] (1.5,0) -- (2.5,1.2) node[right,yshift = 2 pt]{$\pi^+$};
\filldraw [cyan] (1.5,0) circle (3pt);
\end{scope}
\end{tikzpicture}
\caption{ The Feynman diagram contributing to two-body neutron decays $n\to \ell^- \pi^+$, where $\ell =e, \mu$. }
\label{fig:Feyndiagram2b}
\end{figure}
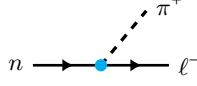

For the two-body neutron decay modes $n\to \ell^- \pi^+$, the leading contribution is mediated by the contact diagram in \cref{fig:Feyndiagram2b}. By evaluating the diagram based on the Lagrangian in \cref{eq:Lnlpi}, we obtain the following general expression for the decay width,
\begin{align}
\label{eq:GammaN2lM}
\Gamma_{n\to \ell^-\pi^+} =&\, \frac{\lambda^{\frac{1}{2}}(m_n^2,m_\ell^2, m_{\pi^+}^2)} {16\pi F_0^2 m_n^3 }
\big[ (m_n^2 + m_\ell^2 -m_{\pi^+}^2)
\nonumber\\
&\times ( |\tilde C_\tR|^2  + |\tilde C_\tL|^2) 
+ 4 m_n m_\ell \Re(\tilde C_\tR \tilde C_{\tL}^*) \big],
\end{align}
where $\lambda(x,y,z)= x^2+y^2+z^2-2xy-2yz-2zx$ is the triangle function.
$\tilde C_{\tR,\tL}$ are related to the LEFT WCs and particle masses in the following manner:
\begin{subequations}
\begin{align}
\tilde C_\tR
=\,& \frac{c_{3\chi}}{4}[2 (m_n^2+m_\ell^2-m_{\pi^+}^2) C_{\partial\bar\ell ddd}^{\tR\tL}
- m_n m_\ell C_{\partial\bar\ell ddd}^{\tL\tR} ]
\nonumber\\
& + c_5 (m_\ell C_{\partial\bar\ell ddd}^{\tR\tR} -m_n C_{\partial\bar\ell ddd}^{\tL\tL}),
\\
\tilde C_\tL  
=\,& \tilde C_\tR\big|_{\tL\leftrightarrow\tR}.
\end{align}
\end{subequations}
Numerically, we obtain 
\begin{subequations}
\label{eq:n2l-pi+}
\begin{align}
\frac{\Gamma_{n\to e^- \pi^+}}{\mathrm{GeV}} =\,& 
0.27 |c_3 C_{\partial\bar eddd}^{\tR\tL}|^2
+1.9 |c_5 C_{\partial\bar eddd}^{\tR\tR}|^2
\nonumber\\
& - 4.1\cdot10^{-4} \Re(c_3 c_5  C_{\partial\bar eddd}^{\tR\tL} C_{\partial\bar eddd}^{\tR\tR*})
\nonumber\\
& +1.5 \cdot10^{-4} c_3^2 C_{\partial\bar eddd}^{\tR\tL} C_{\partial\bar eddd}^{\tL\tR*}
\nonumber\\
& -1.4 \Re(c_3 c_5 C_{\partial\bar eddd}^{\tR\tL} C_{\partial\bar eddd}^{\tL\tL*})
\nonumber\\
& + 4.6\cdot10^{-5} c_5^2  C_{\partial\bar eddd}^{\tR\tR} C_{\partial\bar eddd}^{\tL\tL*}
+\tL\leftrightarrow \tR,
\\
\frac{\Gamma_{n\to \mu^- \pi^+}}{\mathrm{GeV}}=\,& 
0.27 |c_3 C_{\partial\bar\mu ddd}^{\tR\tL}|^2
+1.8 |c_5 C_{\partial\bar\mu ddd}^{\tR\tR}|^2
\nonumber\\
& - 0.085 \Re(c_3 c_5  C_{\partial\bar\mu ddd}^{\tR\tL} C_{\partial\bar\mu ddd}^{\tR\tR*})
\nonumber\\
& + 0.032 c_3^2 C_{\partial\bar\mu ddd}^{\tR\tL} C_{\partial\bar\mu ddd}^{\tL\tR*}
\nonumber\\
& -1.4 \Re(c_3 c_5 C_{\partial\bar\mu ddd}^{\tR\tL} C_{\partial\bar\mu ddd}^{\tL\tL*})
\nonumber\\
& + 0.0093 c_5^2  C_{\partial\bar\mu ddd}^{\tR\tR} C_{\partial\bar\mu ddd}^{\tL\tL*}
+\tL\leftrightarrow \tR,
\end{align}
\end{subequations}
where the unknown LECs are kept explicitly for generality. 

\subsection{Three-body decays $p\to \ell^- \pi^+\pi^+$ and $n\to \ell^- \pi^+(\pi^0,\eta)$ }

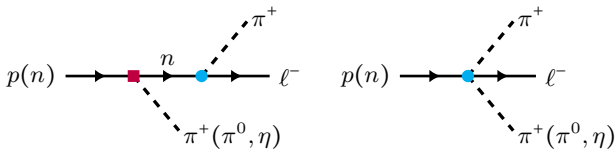
\begin{figure}[ht]
\centering
\begin{tikzpicture}[mystyle,scale=1]
\begin{scope}[shift={(1,1.2)}] 
\draw[f] (0, 0)node[left]{$p(n)$} -- (1.5,0);
\draw[f] (1.5, 0) -- (3,0) node[midway,yshift = 6 pt]{$n$};
\draw[snar, black] (1.5,0) -- (2.5,-1.2) node[right,xshift = -2pt, yshift = -2 pt]{$\pi^+(\pi^0,\eta)$};
\draw[f] (3.0, 0) -- (4.5,0) node[right]{$\ell^-$};
\draw[snar, black] (3,0) -- (4,1.2) node[right,xshift = -2pt, yshift = 2 pt]{$\pi^+$};
\filldraw [purple] (1.4,-0.1) rectangle(1.6,0.1);
\filldraw [cyan] (3,0) circle (3pt);
\end{scope}
\end{tikzpicture}
\hspace{0.2cm}%
\begin{tikzpicture}[mystyle,scale=1]
\begin{scope}[shift={(1,1.2)}] 
\draw[f] (0, 0)node[left]{$p(n)$} -- (1.5,0);
\draw[f] (1.5, 0) -- (3,0) node[right]{$\ell^-$};
\draw[snar, black] (1.5,0) -- (2.5,1.2) node[right,xshift = -2pt, yshift = 2 pt]{$\pi^+$};
\draw[snar, black] (1.5,0) -- (2.5,-1.2) node[right,xshift = -2pt, yshift = -2 pt]{$\pi^+(\pi^0,\eta)$};
\filldraw [cyan] (1.5,0) circle (3pt);
\end{scope}
\end{tikzpicture}
\caption{Leading-order Feynman diagrams for the noncontact (left) and contact (right) contributions to the three-body nucleon decays $p\to \ell^- \pi^+\pi^+$ and $n\to \ell^- \pi^+(\pi^0,\eta)$. The blob and square represent the insertion of a dim-7 BNV and the SM strong interaction vertices, respectively. }
\label{fig:Feyndiagram}
\end{figure}

We next consider the three-body decay modes, ${\tt N}\to \ell^- \pi^+ \bar M~(\tN=p,\,n,~\ell=e,\,\mu,~\bar M=\pi^+,\,\pi^0,\,\eta)$, whose leading-order Feynman diagrams are shown in \cref{fig:Feyndiagram}. 
For notational convenience, we use $\bar M$ to denote the outgoing physical meson, while $M$ denotes the corresponding meson field used below.
In general, the interactions entering these decay vertices for $\tN\to \ell^-\pi^+\bar M$ can be parametrized as follows:
\begin{align}
 {\cal L} &=
\frac{C_{\tN nM}}{F_0} 
\overline{n}\gamma_\mu \gamma_5 \tN \, \partial^\mu M 
\nonumber\\
& - \frac{i}{F_0} \big(C_{n\ell\pi}^{a,\tR}  \pi^- \overline{\partial^\mu\ell_\tL} \tilde{\partial}_\mu {n}_\tR
+ C_{n\ell\pi}^{b,\tR} i \partial_\mu \pi^- \overline{\ell_\tR}\gamma^\mu n_\tR
+ \tL\leftrightarrow \tR\big)
\nonumber\\
& +\frac{1}{F_0^2(1 + \delta_{M,\pi^-})}\Big[ C_{\tN\ell\pi M}^{a,\tR}M \pi^- \overline{\partial^\mu\ell_\tL} \tilde{\partial}_\mu {\tN}_\tR
\label{eq:LB2lM}\\
& + ( C_{\tN\ell\pi M}^{b1,\tR} M i \partial_\mu \pi^-  + C_{\tN\ell\pi M}^{b2,\tR} \pi^-  i \partial_\mu M) \overline{\ell_\tR}\gamma^\mu \tN_\tR -\tL\leftrightarrow \tR\Big],
\nonumber
\end{align}
where the terms in the first two lines correspond to the noncontact contributions shown in \cref{fig:Feyndiagram}, while those in the last two lines correspond to the contact diagram. 
$C_{n\ell\pi}^{a,\tR} =\sqrt{2} c_{3\chi} C_{\partial\bar\ell ddd}^{\tR\tL}$ and  
$C_{n\ell\pi}^{b,\tR}=\sqrt{2} c_5 C_{\partial\bar\ell ddd}^{\tR\tR}$. 
The explicit expressions for  $C_{\tN n M}$ and $C_{\tN\ell\pi M}^{a/b1/b2,\tR}$ for each three-body mode are collected in \cref{tab:Cexpression}. Their chirality-flipped counterparts can be obtained directly by interchanging the superscripts $\tL \leftrightarrow \tR$.
\begin{table}[t]
\center
\resizebox{\linewidth}{!}{
\renewcommand{\arraystretch}{1.3}
\begin{tabular}{l l l l l}
\hline
Mode 
& $C_{\tN n M}$
& $C_{\tN\ell\pi M}^{a,\tR}$
& $C_{\tN\ell\pi M}^{b1,\tR}$
& $C_{\tN\ell\pi M}^{b2,\tR}$
\\\hline
$p\to \ell^- \pi^+\pi^+$ 
& $ \frac{D+F}{\sqrt{2}}$ 
& $- 2 c_{3\chi}C_{\partial\bar\ell ddd}^{\tR\tL}$
& $ - c_5 C_{\partial\bar\ell ddd}^{\tR\tR} $
& $ - c_5 C_{\partial\bar\ell ddd}^{\tR\tR} $ 
\\
$n\to \ell^- \pi^+\pi^0$
& $- \frac{D+F}{2}$ 
& $ \frac{c_{3\chi}}{\sqrt{2}}C_{\partial\bar\ell ddd}^{\tR\tL}$
& $ \frac{3c_5}{\sqrt{2}} C_{\partial\bar\ell ddd}^{\tR\tR} $
& $ -\sqrt{2}c_5 C_{\partial\bar\ell ddd}^{\tR\tR} $
\\
$n\to \ell^- \pi^+\eta$~~~
&$\frac{3F-D}{2\sqrt{3}}$~~~
& $- \frac{c_{3\chi}}{\sqrt{6}}C_{\partial\bar\ell ddd}^{\tR\tL}$~~~
& $ -\sqrt{\frac{3}{2}} c_5 C_{\partial\bar\ell ddd}^{\tR\tR} $~~
& 0
\\\hline
\end{tabular}}
\caption{Summary of $C$-coefficients in  \cref{eq:LB2lM}  for each three-body decay mode. }
\label{tab:Cexpression}
\end{table}

According to the general parametrization in \cref{eq:LB2lM},
and defining $q_{1,2}=p_{1,2}+k$, 
the decay amplitude for $\tN(p) \to \ell^-(k) \pi^+(p_1) \bar M(p_2)$ takes the form:
\begin{align}
F_0^2 {\cal M}= &\, 
\Big\{ C_{\tN n M}\overline{u(k)} \big[ k\cdot\tilde q_1 ( C_{n\ell \pi}^{a,\tR}  P_\tR +  C_{n\ell \pi}^{a,\tL} P_\tL)  - C_{n\ell\pi}^{b,\tR}\slashed{p}_1 P_\tR
\nonumber\\
&- C_{n\ell\pi}^{b,\tL}\slashed{p}_1 P_\tL\big] 
\frac{-1}{\slashed{q}_1- m_n} \slashed{p}_2 \gamma_5 u(p)
+ \delta_{M,\pi^-} (p_1\leftrightarrow p_2)
\Big\}
\nonumber\\
&
+\overline{u(k)} (
C_{\tN\ell\pi M}^{a,\tR} k \cdot \tilde p P_\tR
- C_{\tN\ell\pi M}^{a,\tL} k \cdot \tilde p P_\tL ) u(p)
\nonumber\\
& - \overline{u(k)} \big[(C_{\tN\ell\pi M}^{b1,\tR} \slashed{p}_1+ C_{\tN\ell\pi M}^{b2,\tR}  \slashed{p}_2) P_\tR
\nonumber\\
& -(C_{\tN\ell\pi M}^{b1,\tL}  \slashed{p}_1+ C_{\tN\ell\pi M}^{b2,\tL} \slashed{p}_2) P_\tL \big] u(p).
\label{eq:amplitude1}
\end{align}
Using the Dirac equations and the properties of gamma matrices, the amplitude in \cref{eq:amplitude1} can be reduced into the following form:
\begin{align}
{\cal M}
=\,&F_0^{-2} \overline{u(k)}\big[
\mathbb D_{{\tN;\ell\pi M}}^{\textsc{sr}} P_\tR
+\mathbb  D_{{\tN;\ell\pi M}}^{\textsc{sl}} P_\tL 
\nonumber\\
&+\mathbb  D_{{\tN;\ell\pi M}}^{\textsc{vr}} \slashed{q} P_\tR
+\mathbb  D_{{\tN;\ell\pi M}}^{\textsc{vl}} \slashed{q} P_\tL 
\big]u(p),
\label{eq:amplitude2}
\end{align}
where $q=p_2-p_1$. 
The $\mathbb D$-parameters are functions of the $C$-coefficients in \cref{eq:amplitude1}.  
We define the invariant masses of pairs of final-state particles as $s=(p_1+p_2)^2$, $t=q_2^2= (p_2+k)^2$, and $u=q_1^2=(p_1+k)^2$, 
their explicit expressions are then given as follows: 
\begin{subequations}
\begin{align}
\mathbb D_{{\tN;\ell\pi M}}^{\textsc{sr}} =\,&
\Big\{ \frac{C_{\tN n M}}{m_n^2 - u}
[ C_{n\ell\pi}^{a,\tR}X_1(u) 
+ C_{n\ell\pi}^{a,\tL}X_2(u) 
+ C_{n\ell\pi}^{b,\tR}X_3(u) 
\nonumber
\\ 
& + C_{n\ell\pi}^{b,\tL}X_4(u)]
+ \delta_{M,\pi^-} (u\leftrightarrow t) \Big\}
\nonumber
\\ 
& +\frac{1}{2}(m_\tN^2 +m_\ell^2 -s) C_{\tN\ell\pi M}^{a,\tR}
+\frac{1}{4}m_\tN m_\ell C_{\tN\ell\pi M}^{a,\tL}
\nonumber
\\ 
& +\frac{1}{2} m_\ell (C_{\tN\ell\pi M}^{b1,\tR} + C_{\tN\ell\pi M}^{b2,\tR})
\nonumber
\\ 
& + \frac{1}{2} m_\tN (C_{\tN\ell\pi M}^{b1,\tL} + C_{\tN\ell\pi M}^{b2,\tL}),
\\%
\mathbb D_{{\tN;\ell\pi M}}^{\textsc{sl}} =\,& 
-\mathbb D_{{\tN;\ell\pi M}}^{\textsc{sr}}\big|_{\tL\leftrightarrow \tR}, 
\\%
\mathbb  D_{{\tN;\ell\pi M}}^{\textsc{vr}} =\,& 
\Big\{ \frac{C_{\tN n M}}{m_n^2 - u}
[ C_{n\ell\pi}^{a,\tR}X_5(u) 
+ C_{n\ell\pi}^{a,\tL}X_6(u) 
\nonumber
\\ 
& + C_{n\ell\pi}^{b,\tR}X_7(u)  
+ C_{n\ell\pi}^{b,\tL}X_8(u)]
- \delta_{M,\pi^-} (u\leftrightarrow t)
\Big\}
\nonumber
\\ 
&+ \frac{1}{2}(C_{\tN\ell\pi M}^{b1,\tR} - C_{\tN\ell\pi M}^{b2,\tR}),
\\%
\mathbb  D_{{\tN;\ell\pi M}}^{\textsc{vl}} =\,&
- \mathbb  D_{{\tN;\ell\pi M}}^{\textsc{vr}}\big|_{\tL\leftrightarrow \tR}.  
\end{align}
\end{subequations}
Here, the $X_i$ are functions of the kinematic variables and are defined as follows:  
\begin{subequations}
\begin{align}
X_1(u) =\, &
\frac{1}{8}[ 2(u +m_\ell^2 - m_{\pi^+}^2) (m_\tN^2 - m_\tN m_n - 2 u )
\nonumber
\\ 
& +m_\ell^2 (m_\tN m_n  + u)],
\\ 
X_2(u)=\, & \frac{m_\ell}{8}[ 2(m_\tN + m_n)(m_{\pi^+}^2-m_\ell^2) 
\nonumber
\\ 
& - m_\tN (m_\tN m_n + u)],
\\ 
X_3(u)=\, & \frac{m_\ell}{2}(m_\tN^2 - u), 
\\ 
X_4(u)=\, &\frac{1}{2}[ m_\tN(u-m_\ell^2)
- m_n (m_\tN^2 - 2u + m_\ell^2)],
\\
X_5(u) =\,& - \frac{m_\ell}{8} (m_\tN m_n + u), 
\\
X_6(u) =\,& \frac{1}{4} (m_\tN + m_n )(u + m_\ell^2 - m_{\pi^+}^2),
 \\
X_7(u)=\,& - \frac{1}{2}(m_\tN m_n + u), 
\\
X_8(u)=\,&  \frac{m_\ell}{2}(m_\tN + m_n). 
\end{align}
\end{subequations}
Note that $X_8(u)$ is independent of the variable $u$.

Using \cref{eq:amplitude2}, the spin-averaged and -summed matrix element squared can be calculated straightforwardly and the final expression is provided in Eq.\,(3.5) of \cite{Fan:2026csl}.
In terms of our current notation, it takes the form:
\begin{align}
&F_0^4\overline{|{\cal M}_{\tN\to \ell^-\pi^+ M}|^2} 
\nonumber\\
=\,&  \frac{1}{2}(m_\tN^2 + m_\ell^2 - s)
(|\mathbb D_{{\tN;\ell\pi M}}^{\textsc{sr}}|^2 + |\mathbb D_{{\tN;\ell\pi M}}^{\textsc{sl}}|^2)
\nonumber
\\
& + \frac{1}{2} \big[(m_\tN^2 +m_\ell^2 - s)(s-2m_{\pi^+}^2 - 2 m_M^2) 
+(t-u)^2
\nonumber\\
& - (m_{\pi^+}^2-m_M^2)^2\big] 
(|\mathbb D_{{\tN;\ell\pi M}}^{\textsc{vr}}|^2 
+ |\mathbb D_{{\tN;\ell\pi M}}^{\textsc{vl}}|^2)
\nonumber\\
& + 2 m_\ell m_\tN 
\Re(\mathbb D_{{\tN;\ell\pi M}}^{\textsc{sr}}
\mathbb D_{{\tN;\ell\pi M}}^{\textsc{sl}\,*})
\label{eq:Msquared}
\\
&- 2 m_\ell m_\tN 
(s-2m_{\pi^+}^2 - 2 m_M^2)
\Re(\mathbb D_{{\tN;\ell\pi M}}^{\textsc{vr}}
\mathbb D_{{\tN;\ell\pi M}}^{\textsc{vl}\,*})
\nonumber\\
& + m_\ell(t-u-m_{\pi^+}^2 +m_M^2)
\Re(\mathbb D_{{\tN;\ell\pi M}}^{\textsc{sr}}
\mathbb D_{{\tN;\ell\pi M}}^{\textsc{vr}\,*} 
+ \tL\leftrightarrow \tR)
\nonumber\\
& +m_\tN (t-u+m_{\pi^+}^2-m_M^2) 
\Re(\mathbb D_{{\tN;\ell\pi M}}^{\textsc{sr}}
\mathbb D_{{\tN;\ell\pi M}}^{\textsc{vl}\,*} 
+ \tL\leftrightarrow \tR).
\nonumber
\end{align}
Finally, the total decay width becomes, 
\begin{align}
\Gamma=
\frac{1}{1+\delta_{M,\pi^-}}
\frac{1}{256\pi^3 m_\tN^3}  \int_{s_-}^{s_+} d s \int_{t_-}^{t_+} d t\, 
\overline{|{\cal M}|^2},
\end{align}
where the integration limits are
\begin{align}
s_- & = (m_{\pi^+} +m_M)^2,\quad 
s_+ = (m_\tN - m_\ell)^2,
\nonumber\\
t_\pm & = (E_2^* + E_3^*)^2 - \Big(\sqrt{E_2^{*2} - m_M^2} \mp \sqrt{E_3^{*2} - m_\ell^2} \Big)^2,
\nonumber\\
E_2^* & = 
\frac{s - m_{\pi^+}^2 + m_M^2}{2\sqrt{s} },\quad
E_3^* = 
\frac{ m_\tN^2 - s - m_\ell^2}{2\sqrt{s}}.
\end{align}

Exploiting the above formalism,
we calculate the decay widths for the six three-body nucleon decay modes as functions of the dim-7 LEFT WCs. We use the central values of particle masses and pion decay constant from the Particle Data Group~\cite{ParticleDataGroup:2024cfk}, and
the LECs $D$ and $F$ provided in~\cite{Bali:2022qja}. 
As in the two-body modes $n\to \ell^- \pi^+$ given in \cref{eq:n2l-pi+}, we explicitly keep the $c_{3,5}$ dependence.
The final expressions are 
\begin{subequations}
\begin{align}
\frac{\Gamma_{p\to e^- \pi^+\pi^+}}{10^{-2}\mathrm{GeV}} =\,& 
2.8 |c_3 C_{\partial\bar eddd}^{\tR\tL}|^2
+8.8 |c_5 C_{\partial\bar eddd}^{\tR\tR}|^2
\nonumber\\
& - 0.0076 \Re(c_3 c_5  C_{\partial\bar eddd}^{\tR\tL} C_{\partial\bar eddd}^{\tR\tR*})
\nonumber\\
& - 0.001 c_3^2 C_{\partial\bar eddd}^{\tR\tL} C_{\partial\bar eddd}^{\tL\tR*}
\nonumber\\
& + 6.1 \Re(c_3 c_5 C_{\partial\bar eddd}^{\tR\tL} C_{\partial\bar eddd}^{\tL\tL*})
\nonumber\\
& - 0.017 c_5^2  C_{\partial\bar eddd}^{\tR\tR} C_{\partial\bar eddd}^{\tL\tL*}
+\tL\leftrightarrow \tR,
\\%
\frac{\Gamma_{p\to \mu^- \pi^+\pi^+}}{10^{-2}\mathrm{GeV}} =\,& 
2.7 |c_3 C_{\partial\bar\mu ddd}^{\tR\tL}|^2
+7.6 |c_5 C_{\partial\bar\mu ddd}^{\tR\tR}|^2
\nonumber\\
& -1.4 \Re(c_3 c_5  C_{\partial\bar\mu ddd}^{\tR\tL} C_{\partial\bar\mu ddd}^{\tR\tR*})
\nonumber\\
& -0.21 c_3^2 C_{\partial\bar\mu ddd}^{\tR\tL} C_{\partial\bar\mu ddd}^{\tL\tR*}
\nonumber\\
& +5.7 \Re(c_3 c_5 C_{\partial\bar\mu ddd}^{\tR\tL} C_{\partial\bar\mu ddd}^{\tL\tL*})
\nonumber\\
& -2.6 c_5^2  C_{\partial\bar\mu ddd}^{\tR\tR} C_{\partial\bar\mu ddd}^{\tL\tL*}
+\tL\leftrightarrow \tR,
\\%
\frac{\Gamma_{n\to e^- \pi^+\pi^0}}{10^{-2}\mathrm{GeV}}  =\,& 
0.79 |c_3 C_{\partial\bar eddd}^{\tR\tL}|^2
+29 |c_5 C_{\partial\bar eddd}^{\tR\tR}|^2
\nonumber\\
& - 0.0042 \Re(c_3 c_5  C_{\partial\bar eddd}^{\tR\tL} C_{\partial\bar eddd}^{\tR\tR*})
\nonumber\\
& -2.1\cdot10^{-4} c_3^2 C_{\partial\bar eddd}^{\tR\tL} C_{\partial\bar eddd}^{\tL\tR*}
\nonumber\\
& -0.55 \Re(c_3 c_5 C_{\partial\bar eddd}^{\tR\tL} C_{\partial\bar eddd}^{\tL\tL*})
\nonumber\\
& +0.034 c_5^2  C_{\partial\bar eddd}^{\tR\tR} C_{\partial\bar eddd}^{\tL\tL*}
+\tL\leftrightarrow \tR,
\\%
\frac{\Gamma_{n\to \mu^- \pi^+\pi^0}}{10^{-2}\mathrm{GeV}} =\,& 
0.78 |c_3 C_{\partial\bar\mu ddd}^{\tR\tL}|^2
+25 |c_5 C_{\partial\bar\mu ddd}^{\tR\tR}|^2
\nonumber\\
& -0.83 \Re(c_3 c_5  C_{\partial\bar\mu ddd}^{\tR\tL} C_{\partial\bar\mu ddd}^{\tR\tR*})
\nonumber\\
& -0.043 c_3^2 C_{\partial\bar\mu ddd}^{\tR\tL} C_{\partial\bar\mu ddd}^{\tL\tR*}
\nonumber\\
& -0.56 \Re(c_3 c_5 C_{\partial\bar\mu ddd}^{\tR\tL} C_{\partial\bar\mu ddd}^{\tL\tL*})
\nonumber\\
& +5.1 c_5^2  C_{\partial\bar\mu ddd}^{\tR\tR} C_{\partial\bar\mu ddd}^{\tL\tL*}
+\tL\leftrightarrow \tR,
\\%
\frac{\Gamma_{n\to e^- \pi^+\eta}}{10^{-4}\mathrm{GeV}}  =\,& 
0.22 |c_3 C_{\partial\bar eddd}^{\tR\tL}|^2
+13 |c_5 C_{\partial\bar eddd}^{\tR\tR}|^2
\nonumber\\
& - 0.0057 \Re(c_3 c_5  C_{\partial\bar eddd}^{\tR\tL} C_{\partial\bar eddd}^{\tR\tR*})
\nonumber\\
& -3.3\cdot10^{-4} c_3^2 C_{\partial\bar eddd}^{\tR\tL} C_{\partial\bar eddd}^{\tL\tR*}
\nonumber\\
& +2.4 \Re(c_3 c_5 C_{\partial\bar eddd}^{\tR\tL} C_{\partial\bar eddd}^{\tL\tL*})
\nonumber\\
& -0.01 c_5^2  C_{\partial\bar eddd}^{\tR\tR} C_{\partial\bar eddd}^{\tL\tL*}
+\tL\leftrightarrow \tR,
\\%
\frac{\Gamma_{n\to \mu^- \pi^+\eta}}{10^{-4}\mathrm{GeV}}   =\,& 
0.16 |c_3 C_{\partial\bar\mu ddd}^{\tR\tL}|^2
+6.4 |c_5 C_{\partial\bar\mu ddd}^{\tR\tR}|^2
\nonumber\\
& -0.72 \Re(c_3 c_5  C_{\partial\bar\mu ddd}^{\tR\tL} C_{\partial\bar\mu ddd}^{\tR\tR*})
\nonumber\\
& -0.056 c_3^2 C_{\partial\bar\mu ddd}^{\tR\tL} C_{\partial\bar\mu ddd}^{\tL\tR*}
\nonumber\\
& +1.6 \Re(c_3 c_5 C_{\partial\bar\mu ddd}^{\tR\tL} C_{\partial\bar\mu ddd}^{\tL\tL*})
\nonumber\\
& -1.1 c_5^2  C_{\partial\bar\mu ddd}^{\tR\tR} C_{\partial\bar\mu ddd}^{\tL\tL*}
+\tL\leftrightarrow \tR.
\end{align}
\end{subequations}

Compared to the single-pion modes in \cref{eq:n2l-pi+}, the decay widths of the two-pion modes are generally suppressed by factors of $10^{-2}\text{--}10^{-1}$,  with the precise suppression depending on the specific decay mode and the operator involved. 
The $\eta$-related modes are further suppressed by factors of $10^{-5}\text{--}10^{-3}$, due to their more limited phase space. 

In the above calculation, we have neglected contributions mediated by vector mesons, which were estimated in \cite{Fan:2026fqo,Fan:2026csl} to be less than $30\,\%$ for similar three-body decay modes involving $\pi^+\pi^0$ final states. 
We have also neglected hadronic uncertainties and higher-order chiral corrections, which we expect to result in corrections at most at the $\calO(1)$ level. 
A more detailed treatment of these effects is left for future work.

\subsection{Matching onto dim-7 BNV SMEFT operators}

Having obtained the LEFT results, we next establish their leading-order SMEFT counterparts.
This can be accomplished by identifying the relevant SMEFT interactions and matching them onto the corresponding LEFT WCs. 
Owing to the particular pattern of global baryon- and lepton-number violation in the SMEFT, we find that the relevant leading-order interactions first arise at dimension 7. 
There are two such derivative-type operators~\cite{Lehman:2014jma,Liao:2016hru}: 
\begin{subequations}
\label{eq:SMEFTdim7ope}
\begin{align}
\calO^{prst}_{\bar{L} QddD}=\,& \epsilon_{\alpha\beta\gamma}
(\overline{L_{p}} \gamma_\mu Q^{\alpha}_{r})
(\overline{d^{\beta \C}_s} i D^\mu d^{\gamma}_t ),
\\
\calO^{prst}_{\bar{e}dddD} =\, &
\epsilon_{\alpha \beta\gamma}
(\overline{e_{p}} \gamma_\mu d^{\alpha}_r) (\overline{d^{\beta \C}_s}i D^\mu d^{\gamma}_t),
\end{align}
\end{subequations}
where $L$ and $Q$ ($e$ and $d$) denote the SM lepton and quark doublets (singlets), respectively. 

The matching from the SMEFT onto the LEFT depends on the choice of quark flavor-to-mass rotations. Two commonly used choices are the down-quark and up-quark flavor bases. In these bases,
the Cabibbo-Kobayashi-Maskawa (CKM) matrix $V$ enters the left-handed quark rotations according to $u_\tL'=V^\dagger u_\tL$ in the down-quark basis and $d'_{\tL}=V d_{\tL}$ in the up-quark basis,  where the primed fields denote the flavor eigenstates.
In the up-quark flavor basis,
we obtain the following nonvanishing LEFT matching results responsible for these $\Delta I=3/2$ transitions~\cite{Liao:2026ugl},
\begin{align}
C_{\partial \bar e(\bar\mu) ddd}^{\tR\tL} =\,& - 0.91\times V_{w1}C_{\bar LQddD}^{1(2)w11},
\nonumber\\
C_{\partial \bar e(\bar\mu) ddd}^{\tR\tR} =\,& 0.76\times \frac{1}{2} C_{\bar edddD}^{1(2)111}. 
\label{eq:matching}
\end{align} 
We note that the corresponding chirality-flipped partners cannot be generated at the dim-7 level; 
their first nontrivial contributions instead arise from dim-9 or higher SMEFT operators.
In the expressions above, we have included the QCD running factors, 0.91 and 0.76, associated with the RG evolution from the electroweak scale down to the chiral scale $\Lambda_\chi$. 
The same matching results apply in the down-quark flavor basis upon setting the CKM matrix element to the identity and the generation label $w=1$. 

Combining the matching results in \cref{eq:matching}, the resulting decay widths expressed in terms of the dim-7 SMEFT WCs are summarized below:
\begin{subequations}
\label{eq:decay_width_smeft}
\begin{align}
\frac{\Gamma_{n\to e^- \pi^+}}{\mathrm{GeV}}  =\,& 
0.23 |c_3 V_{w1}C_{\bar LQddD}^{1w11}|^2
+0.27 |c_5 C_{\bar edddD}^{1111}|^2
\nonumber\\
+&  1.4\cdot10^{-4} \Re(c_3 c_5  V_{w1}C_{\bar LQddD}^{1w11} C_{\bar edddD}^{1111*}), 
\\
\frac{\Gamma_{n\to \mu^- \pi^+}}{\mathrm{GeV}} =\,& 
0.23 |c_3 V_{w1} C_{\bar LQddD}^{2w11}|^2
+0.26 |c_5 C_{\bar edddD}^{2111}|^2
\nonumber\\
& + 0.029\Re(c_3 c_5  V_{w1}C_{\bar LQddD}^{2w11} C_{\bar edddD}^{2111*}), 
\\%
\frac{\Gamma_{p\to e^- \pi^+\pi^+} }{10^{-2}\mathrm{GeV}} =\,& 
2.3 |c_3 V_{w1}C_{\bar LQddD}^{1w11}|^2
+1.3 |c_5 C_{\bar edddD}^{1111}|^2
\nonumber\\
& + 0.0026 \Re(c_3 c_5  V_{w1}C_{\bar LQddD}^{1w11} C_{\bar edddD}^{1111*}), 
\\
\frac{\Gamma_{p\to \mu^- \pi^+\pi^+} }{10^{-2}\mathrm{GeV}} =\,& 
2.3 |c_3 V_{w1} C_{\bar LQddD}^{2w11}|^2
+1.1 |c_5 C_{\bar edddD}^{2111}|^2
\nonumber\\
& + 0.50\Re(c_3 c_5  V_{w1}C_{\bar LQddD}^{2w11} C_{\bar edddD}^{2111*}), 
\\%
\frac{\Gamma_{n\to e^- \pi^+\pi^0} }{10^{-2}\mathrm{GeV}}  =\,& 
0.66 |c_3 V_{w1}C_{\bar LQddD}^{1w11}|^2
+4.1 |c_5 C_{\bar edddD}^{1111}|^2
\nonumber\\
& + 0.0015 \Re(c_3 c_5  V_{w1}C_{\bar LQddD}^{1w11} C_{\bar edddD}^{1111*}), 
\\
\frac{\Gamma_{n\to \mu^- \pi^+\pi^0} }{10^{-2}\mathrm{GeV}} =\,& 
0.64 |c_3 V_{w1} C_{\bar LQddD}^{2w11}|^2
+3.6 |c_5 C_{\bar edddD}^{2111}|^2
\nonumber\\
& + 0.29 \Re(c_3 c_5  V_{w1}C_{\bar LQddD}^{2w11} C_{\bar edddD}^{2111*}), 
\\%
\frac{\Gamma_{n\to e^- \pi^+\eta} }{10^{-4}\mathrm{GeV}} =\,& 
0.18 |c_3 V_{w1}C_{\bar LQddD}^{1w11}|^2
+1.9 |c_5 C_{\bar edddD}^{1111}|^2
\nonumber\\
& + 0.0020 \Re(c_3 c_5  V_{w1}C_{\bar LQddD}^{1w11} C_{\bar edddD}^{1111*}), 
\\
\frac{\Gamma_{n\to \mu^- \pi^+\eta} }{10^{-4}\mathrm{GeV}}  =\,& 
0.13 |c_3 V_{w1} C_{\bar LQddD}^{2w11}|^2
+0.93 |c_5 C_{\bar edddD}^{2111}|^2
\nonumber\\
& + 0.25 \Re(c_3 c_5  V_{w1}C_{\bar LQddD}^{2w11} C_{\bar edddD}^{2111*}).
\end{align}
\end{subequations}

\begin{figure}[t]
\centering
\includegraphics[width=0.238\textwidth]{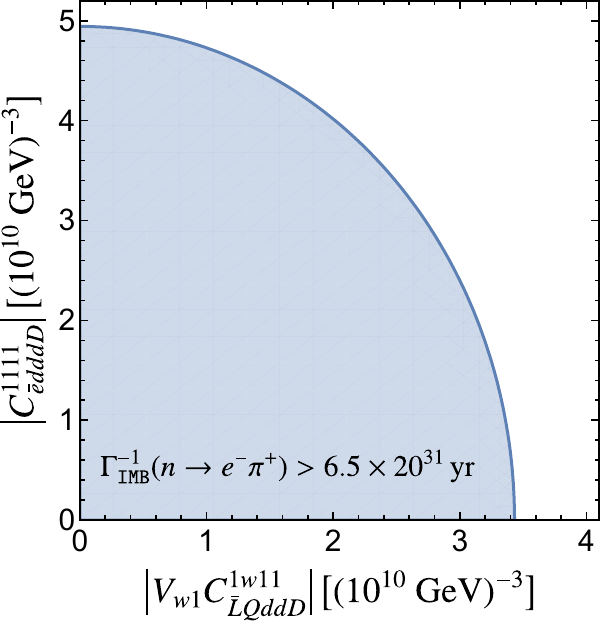}
\includegraphics[width=0.238\textwidth]{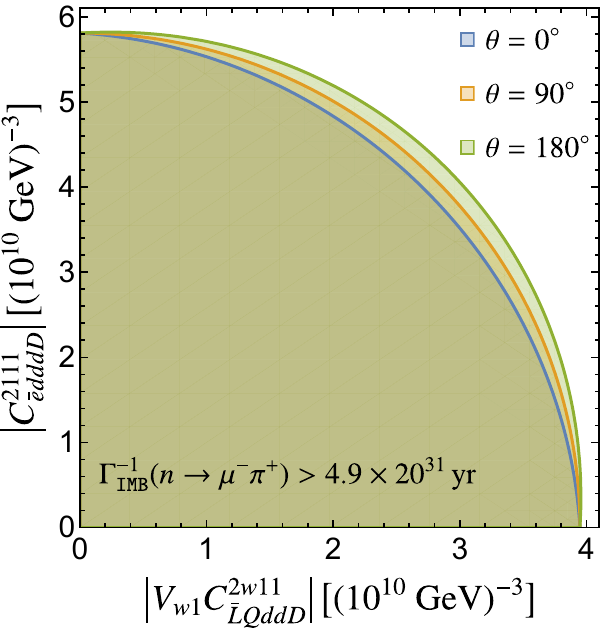}
\caption{Allowed regions in the WC-magnitude plane derived from the existing IMB bounds on the two-body decays $n\to e^-\pi^+$ (left panel) and $n\to \mu^-\pi^+$ (right panel).}
\label{fig:WCsbound}    
\end{figure}

We now investigate constraints on the two SMEFT WC combinations using the existing 90\,\%~C.L. lower bounds on the partial lifetimes of the two two-body decay modes from the IMB experiment~\cite{Seidel:1988ut}, 
\begin{subequations}
\begin{align}
\Gamma^{-1}_{\tt IMB}(n\to e^-\pi^+)
\gtrsim&\, 6.5\times 10^{31}~\mathrm{yr},
\\
\Gamma^{-1}_{\tt IMB}(n\to \mu^-\pi^+)
\gtrsim&\, 4.9\times 10^{31}~\mathrm{yr}.
\end{align}
\end{subequations}
Since each decay width depends only on the magnitudes of the two WCs $|V_{w1} C_{\bar LQddD}^{x w11}|$ and $|C_{\bar edddD}^{x 111}|$ and their relative phase $\theta=\arg(V_{w1} C_{\bar LQddD}^{x w11}C_{\bar edddD}^{x 111*})$,
we present the resulting constraints in the two-dimensional plane spanned by $|V_{w1} C_{\bar LQddD}^{x w11}|$ and $|C_{\bar edddD}^{x 111}|$ for several representative values of $\theta$.
For this analysis, we fix the LECs $c_{3,5}$ to their NDA values, and the resulting allowed regions are shown in \cref{fig:WCsbound}. 
The left and right panels correspond to the electron and muon flavors, respectively.
Note that the constraint shown in the left panel is insensitive to the value of $\theta$, reflecting the fact that the interference term in the electron mode $n\to e^-\pi^+$ is negligible. These constraints correspond to effective scales of up to approximately $5\times 10^9~\rm GeV$ , illustrating the high sensitivity of nucleon decay to very heavy new physics. 

\begin{figure}[t]
\centering
\includegraphics[width=0.48\textwidth]{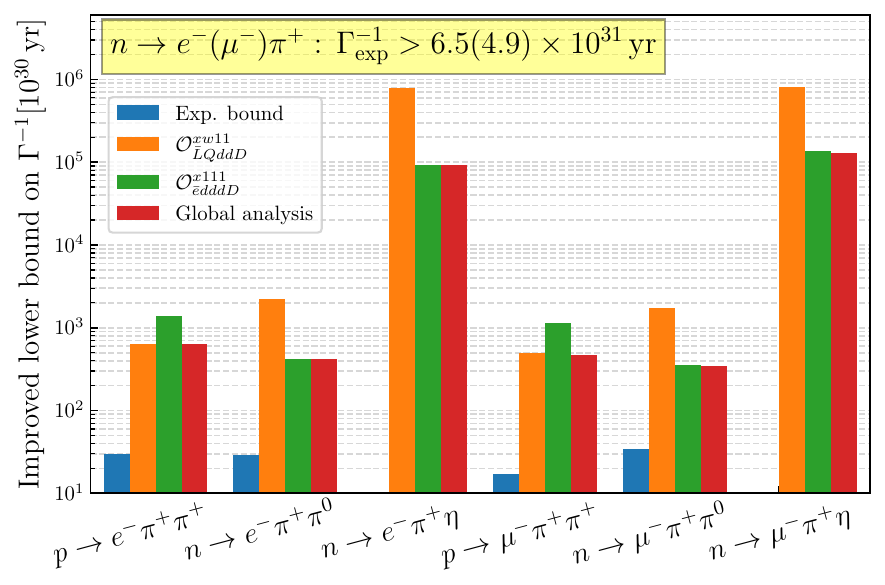}
\caption{The improved lower partial lifetime limits for three-body nucleon decay modes. The lepton generation $x=1(2)$ for electron (muon) flavor. The experimental bounds for the two-pion modes are taken from~\cite{Frejus:1991ben}.}
\label{fig:newInGbound}    
\end{figure}

Since the three-body modes are correlated with the two two-body modes through the same set of WCs, the experimental bounds on the latter can be used to derive more stringent bounds on the former. 
We consider two approaches: a single-operator-dominance analysis and a global analysis in which the two WC combinations are simultaneously taken into account.
In the first approach, we assume that only one of the two relevant WCs is nonzero at a time.
Each three-body decay width is then directly proportional to that of the corresponding two-body mode with the same lepton flavor, allowing the experimental bound on the latter to be directly translated into a bound on the former. The resulting operator-dependent lower limits on the partial lifetimes are shown by the orange and green bars in \cref{fig:newInGbound}. 

For the global analysis, we simultaneously consider both WCs and derive conservative lower bounds on the three-body modes using algebraic inequalities.   
To this end, we define $z_1= |c_3 V_{w1} C_{\bar LQddD}^{x w11}|>0$ and $z_2= |c_5 C_{\bar edddD}^{x 111}|>0$. 
For each of the four decay modes associated with the electron or muon flavor in \cref{eq:decay_width_smeft}, the decay width can be written in a common form $\Gamma= a z_1^2 + b z_2^2 + c \cos\theta\, z_1 z_2$, where $a,b,c$ are positive numerical prefactors that differ among the four modes for each lepton flavor.  
We then obtain the following relation between the partial lifetimes of the three- to two-body decay modes:
\begin{align}
\Gamma^{-1}_{\rm 3B} =\, &
\frac{a_{\rm 2B} z_1^2 + b_{\rm 2B} z_2^2 + c_{\rm 2B} \cos\theta\, z_1 z_2}{a_{\rm 3B} z_1^2 + b_{\rm 3B} z_2^2 + c_{\rm 3B} \cos\theta\, z_1 z_2}
\Gamma^{-1}_{\rm 2B}  
\nonumber\\
\gtrsim\,&\frac{a_{\rm 2B} z_1^2 + b_{\rm 2B} z_2^2 - c_{\rm 2B} z_1 z_2}{a_{\rm 3B} z_1^2 + b_{\rm 3B} z_2^2 + c_{\rm 3B} z_1 z_2}\Gamma^{-1,\,90\,\%}_{\rm 2B,\,\tt IMB}
\nonumber\\
=\,&\frac{a_{\rm 2B} + b_{\rm 2B} z^2 - c_{\rm 2B} z}{a_{\rm 3B} + b_{\rm 3B} z^2 + c_{\rm 3B} z}\Gamma^{-1,\,90\,\%}_{\rm 2B,\,\tt IMB}, 
\end{align}
where $z= z_2/z_1\gtrsim0$. 
The subscripts 2B and 3B denote the corresponding parameters for the two-body and three-body modes, respectively. 
In the second step, we used the fact that the numerator (denominator) is bounded from below (above) by its value at $\cos\theta=-1 (+1)$.
The minimum of the function in the last line provides a conservative lower limit on the partial lifetime of each three-body mode. The final bounds obtained in this way are shown by red bars in \cref{fig:newInGbound}. 
Although these bounds are weaker than the operator-dependent bounds, they are more robust, as they do not rely on the single-operator-dominance assumption. 

In \cref{fig:newInGbound}, we also show the previous Frejus bounds on the partial lifetimes of two-pion modes by blue bars~\cite{Frejus:1991ben}, which are at the level of $(2\text{--}3)\times10^{31}\,\rm yr$.
We find that our indirect EFT-consistency bounds are stronger than the Frejus results by one to two orders of magnitude. For the two $\eta$-related modes, no experimental bounds are currently available, while our indirect bounds reach approximately $\calO(10^{35}\,\rm yr)$. Moreover, if future experiments improve the sensitivity to $n\to (e^-,\mu^-)\pi^+$ by one order of magnitude relative to the current IMB limits, the corresponding indirect bounds for the three-body modes would also improve by the same order of magnitude.

\section{A UV model}
\label{sec:uvmodels}

To connect our formalism to new physics scenarios, we present a renormalizable UV model that generates the dim-7 BNV SMEFT operator $\calO_{\bar edddD}$ at leading order while forbidding the dim-6 BNV interactions. 
We introduce three heavy new fields beyond the SM:
a right-handed sterile fermion $N_\tR(\mathbf{1},\mathbf{1},0)_{-1}$, 
a scalar leptoquark $S(\mathbf{3},\mathbf{1},-1/3)_{-1}$, 
and a vector-like fermion $U(\mathbf{3},\mathbf{1},2/3)_{-1}$, where the numbers in brackets refer to their representations under the SM gauge group $\rm SU(3)_c\otimes SU(2)_L\otimes U(1)_Y$. 
We further impose a discrete $\mathbb Z_2$ symmetry under which the three heavy fields are odd and all SM particles are even. 
Denoting the SM right-handed charged-lepton and down-type quark fields by $e_p$ and $d_p$, with $p=1,2,3$ a generation index, 
the most general $\mathbb Z_2$-invariant Lagrangian is
\begin{align}
\mathcal{L}_{\rm NP} =
\,&
\mathcal{L}_{\rm SM}+ \overline{U} i\slashed{D}U - M_U \overline{U}U 
\nonumber\\
& + \overline{N_\tR} i \slashed{D} N_\tR- \frac{1}{2} ( M_N \overline{N_\tR^\C}N_\tR 
+\mathrm{H.c.} )
\nonumber\\
& + |D_\mu S|^2- M_S^2 S^\dagger S
 - \lambda_{HS}\,S^\dagger S\, H^\dagger H
- \lambda_{S} |S^\dagger S|^2
\nonumber\\
&- \big[ y_{1,p} (\overline{e_{p}} U_\tR^\C) S
+ y_{2,p}\epsilon_{\alpha\beta\gamma} (\overline{U_\tR^{\alpha\C}} d_{p}^\beta)S^\gamma
 \nonumber\\
&  + y_{3,p} S^\dagger(\overline{N_\tR^\C} d_{p})
+\mathrm{H.c.} \big].
\end{align}
Here, $\mathcal{L}_{\rm SM}$ represents the standard SM Lagrangian. 
The masses of $N_\tR$, $S$, and $U$ are denoted by
$M_N$, $M_S$, and $M_U$, respectively,
and $N_\tR^\C$ and $U^\C$ denote their 
charge-conjugation fields. 
$y_{1/2/3,p}$ represent the new Yukawa couplings, while $\lambda_{HS}$ and $\lambda_S$ are new Higgs quartic couplings.

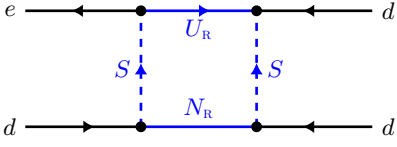
\begin{figure}[t]
\centering
\begin{tikzpicture}[mystyle,scale=1.7]
\begin{scope}
\draw[f] (1.5, 0) -- (0,0)node[left]{$e$};
\draw[f, blue] (1.5, 0) -- (3,0) node[midway,yshift = -7 pt]{$U_\tR$} ;
\draw[f] (4.5, 0) node[right]{$d$} -- (3,0);
\draw[s, blue] (1.5,-1.5)--(1.5,0)   node[midway,xshift = -7 pt]{$S$};
\draw[s, blue] (3,-1.5)-- (3,0)  node[midway,xshift = 7 pt]{$S$};
\draw[f] (0, -1.5)node[left]{$d$} -- (1.5,-1.5);
\draw[blue] (1.5, -1.5) -- (3,-1.5) node[midway,yshift = 7 pt]{$N_\tR$} ;
\draw[f] (4.5, -1.5) node[right]{$d$} -- (3,-1.5);
\filldraw [black] (1.5,0) circle (1.5pt);
\filldraw [black] (3,0) circle (1.5pt);
\filldraw [black] (1.5,-1.5) circle (1.5pt);
\filldraw [black] (3,-1.5) circle (1.5pt);
\end{scope}
\end{tikzpicture}
\caption{One-loop Feynman diagram responsible for the dim-7 BNV operator $\calO_{\bar edddD}$.}
\label{fig:Feyndiagram_dim7ope}
\end{figure}

Assigning baryon plus lepton number ($B+L$) charges $Q_{B+L}(N_\tR)=0$,
$Q_{B+L}(S)=1/3$, and $Q_{B+L}(U)=-2/3$, all interaction terms in $\mathcal{L}_{\rm NP} $ conserve $B+L$, while $B$ and $L$ are individually violated.
Since the dim-6 and dim-8 BNV SMEFT operators necessarily violate $B+L$, they are forbidden in this model. 
The leading BNV interactions arise at dimension 7 and are generated at one loop through the topology shown in \cref{fig:Feyndiagram_dim7ope}. 
Integrating out the heavy particles using {\tt Matchete} code~\cite{Fuentes-Martin:2022jrf}, we obtain the following effective interaction: 
\begin{align}
{\cal L}_{\slashed{B},\rm eff} = \sum_{p,r,s,t} \calO_{\bar edddD}^{prst} C_{\bar edddD}^{prst} +\mathrm{H.c.}, 
\end{align}
where $\calO_{\bar edddD}^{prst}$ is defined in \cref{eq:SMEFTdim7ope}, and the corresponding matched WC takes 
\begin{align}
C_{\bar edddD}^{prst}  =\frac{y_{1,p}\, y_{2,r}\, y_{3,s}\, y_{3,t}}{32\pi^2 M_N^3}  L(r_1,r_2).
\end{align}
Denoting the mass ratios $ r_1 = M_S^2/M_N^2$ and $r_2= M_U^2/M_N^2$, the loop function $L(r_1,r_2)$ takes the form
\begin{align}
L(r_1,r_2) =&\, \frac{r_1( r_1 -2 r_2 +r_1^2)\ln r_1 }{(1-r_1)^3 (r_1-r_2)^2}
+\frac{r_2^2 \ln r_2 }{(1-r_2)^2 (r_1-r_2)^2}
\nonumber\\
& + \frac{2 r_1 - r_2 - r_1 r_2 }{(1-r_1)^2(1-r_2)(r_1-r_2)}
\overset{r_{1,2}\to1 }{\Longrightarrow}- \frac{1}{12}. 
\end{align}

\begin{figure}[t]
\centering
\includegraphics[width=0.45\textwidth]{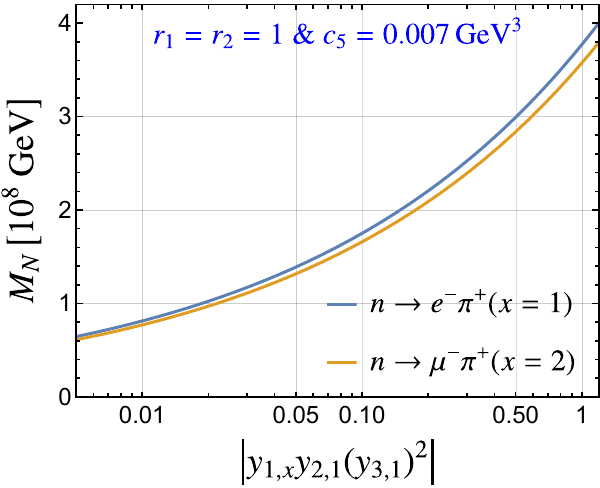}
\caption{The constraint on the NP scale $M_N$ as a function of the Yukawa coupling combination, derived from IMB bounds on the neutron decay modes $n\to (e^-,\mu^-) \pi^+$.}
\label{fig:newbound}    
\end{figure}

We now focus on the case of first-generation quarks, which directly corresponds to the $\Delta I=3/2$ processes discussed above. In this case, the SMEFT WCs in  \cref{eq:decay_width_smeft} reduce to
\begin{align}
C_{\bar edddD}^{x111}
= \frac{y_{1,x}\, y_{2,1}\, (y_{3,1})^2}{32\pi^2 M_N^3}  L(r_1,r_2), 
\label{eq:UV_matching}
\end{align}
while $C_{\bar LQddD}^{xw11}= 0$. 
For the numerical analysis, we consider the degenerate heavy-mass limit, $r_1=r_2=1$, for which the loop function reduces to the constant $-1/12$.  
Substituting \cref{eq:UV_matching} into \cref{eq:decay_width_smeft} and employing the IMB bounds on the two-body decay modes, we obtain constraints on the NP parameters.  
\cref{fig:newbound} shows the corresponding constraint on the NP scale $M_N$ as a function of the Yukawa coupling combination.  
For $\calO(1)$ Yukawa couplings, the NP scale can be probed up to $(1\text{--}4)\times10^8\,\rm GeV$, which is generally lower than the effective scales derived above due to the loop suppression.

Before closing this section, we note that the sterile fermion $N_\tR$ in this model could potentially serve as a viable dark matter candidate if it is the lightest $\mathbb Z_2$-odd particle. Moreover, 
the simultaneous violation of baryon and lepton numbers could provide a framework for generating the observed matter–antimatter asymmetry through CP-violating, out-of-equilibrium dynamics involving the heavy states.
We leave a detailed investigation of these intriguing possibilities to future work.

\section{Summary}
\label{sec:summary}

In this work, we systematically classify the dim-7 BNV LEFT operators involving derivatives and show that they can be organized into two general chiral structures and their chirality-flipped counterparts. Focusing on the operators involving light $u,d,s$ quarks, we match them onto ChPT at leading order to compute the corresponding nucleon decay matrix elements within a systematic and consistent framework, with the associated LECs estimated using NDA. 

We then apply this chiral framework to the experimentally interesting $\Delta I=3/2$ nucleon decays mediated by these derivative-type operators at leading order. These include $n\to \ell^-\pi^+$ and the corresponding three-body modes $p\to\ell^-\pi^+\pi^+$ and $n\to\ell^-\pi^+(\pi^0,\eta)$, with $\ell=e,\mu$. We derive explicit expressions for the decay widths in terms of both LEFT and SMEFT WCs. Using the existing experimental bounds on $n\to \ell^-\pi^+$, we obtain stringent constraints on the relevant SMEFT WCs. 
Furthermore, exploiting the EFT correlations between the two- and three-body decay modes, we derive indirect lower limits on the partial lifetimes of the six three-body modes. The resulting limits reach $\calO(3\times 10^{32}\,\rm yr)$ for $p\to\ell^-\pi^+\pi^+$ and $n\to \ell^-\pi^+\pi^0$, and $\calO(10^{35}\,\rm yr)$ for $n\to\ell^-\pi^+\eta$, substantially exceeding the currently available experimental bounds. These results provide a systematic framework for studying dim-7 contributions to nucleon decays. 

Finally, we construct a UV model that generates the relevant dim-7 SMEFT/LEFT operators and the associated $\Delta I=3/2$ nucleon decay modes at leading order. This provides further motivation for experimental searches for the $\Delta I=3/2$ nucleon decays, as the underlying UV physics responsible for these modes may offer connections to broader open questions, including the nature of dark matter and the origin of the matter-antimatter asymmetry of the Universe.

\acknowledgments
We thank Yi Liao, Jin-Han Liang, and Hao-Lin Wang for useful comments on the manuscript. This work was supported 
by Grants No.\,NSFC-12305110
and No.\,NSFC-12035008.

\appendix
\newlength{\fwidth}
\setlength{\fwidth}{0.3\textwidth}

\section{Demonstration of operator basis equivalence}
\label{app:bases_trans}
By employing the Fierz identity
\begin{align}
(\overline{\Psi_{1\tR}}\gamma^\mu \Psi_{2\tR})(\overline{\Psi_{3\tL}^\C} \Psi_{4\tL})=\,&
-(\overline{\Psi_{1\tR}} \Psi_{3\tL})(\overline{\Psi_{4\tL}^\C}\gamma^\mu \Psi_{2\tR})
\nonumber\\
&-(\overline{\Psi_{1\tR}} \Psi_{4\tL})(\overline{\Psi_{3\tL}^\C}\gamma^\mu \Psi_{2\tR}),
\label{eq:FI}
\end{align}
and the integration by parts (IBPs), we obtain 
\begin{eqnarray}
&&(\overline{\Psi_{1\tR}}\gamma^\mu \Psi_{2\tR})(\overline{\Psi_{3\tL}^\C} i \overleftrightarrow{D}_\mu\Psi_{4\tL})
\nonumber\\
&\overset{\rm(\ref{eq:FI})}{=}&
(\overline{\Psi_{1\tR}} i D_\mu \Psi_{3\tL})(\overline{\Psi_{4\tL}^\C}\gamma^\mu \Psi_{2\tR})
\nonumber\\
&&+ i (\overline{\Psi_{1\tR}} \Psi_{4\tL})(\overline{ D_\mu\Psi_{3\tL}^\C}\gamma^\mu \Psi_{2\tR}) - 3\leftrightarrow 4
\nonumber\\
&\overset{\rm IBP}{=}&
- i(\overline{D_\mu \Psi_{1\tR}}\Psi_{3\tL})(\overline{\Psi_{4\tL}^\C}\gamma^\mu \Psi_{2\tR})
\nonumber\\
&&-i(\overline{\Psi_{1\tR}} \Psi_{3\tL})(\overline{D_\mu\Psi_{4\tL}^\C}\gamma^\mu \Psi_{2\tR})
\nonumber\\
&&-(\overline{\Psi_{1\tR}}\Psi_{3\tL})(\overline{\Psi_{4\tL}^\C}i\slashed{D} \Psi_{2\tR})
\nonumber\\
&&+ i (\overline{\Psi_{1\tR}} \Psi_{4\tL})(\overline{ D_\mu\Psi_{3\tL}^\C}\gamma^\mu \Psi_{2\tR}) - 3\leftrightarrow 4 +\cdots
\nonumber\\
&=&
- i(\overline{D_\mu \Psi_{1\tR}}\Psi_{3\tL})(\overline{\Psi_{4\tL}^\C}\gamma^\mu \Psi_{2\tR})
\nonumber\\
&&+ (\overline{\Psi_{1\tR}} \Psi_{3\tL})(\overline{\Psi_{2\tR}^\C}i\slashed{D} \Psi_{4\tL})
- (\overline{\Psi_{1\tR}} \Psi_{4\tL})(\overline{ \Psi_{2\tR}^\C}i\slashed{D} \Psi_{3\tL})
\nonumber\\
&&-(\overline{\Psi_{1\tR}}\Psi_{3\tL})(\overline{\Psi_{4\tL}^\C}i\slashed{D} \Psi_{2\tR})
- 3\leftrightarrow 4 +\cdots,
\end{eqnarray}
where the ellipsis denotes total derivative terms.
For the $q_\tR q_\tL q_\tL$-type operators with derivatives acting on quark fields, by replacing the general fermion fields by the corresponding lepton and quark fields,  the above transformation gives 
\begin{align}
& (\overline{l_\tR}\gamma^\mu q_{\tR,y}^\alpha)   
(\overline{q_{\tL,z}^{\beta\C}}i\overleftrightarrow{D_\mu}q_{\tL,w}^\gamma)\epsilon_{\alpha\beta\gamma}
\nonumber\\
=\,& - 2 i(\overline{D_\mu l_\tR}q_{\tL,\{z}^\alpha)
(\overline{q_{\tL,w\}}^{\beta\C}}\gamma^\mu q_{\tR,y}^\gamma)\epsilon_{\alpha\beta\gamma}
\nonumber\\
&- 2 (\overline{l_\tR}q_{\tL,\{z}^\alpha)
(\overline{q_{\tL,w\}}^{\beta\C}}i\slashed{D} q_{\tR,y}^\gamma)\epsilon_{\alpha\beta\gamma}
\nonumber\\
&- 4 (\overline{l_\tR}q_{\tL,\{z}^\alpha)
(\overline{q_{\tR,y}^{\beta\C}}i\slashed{D} q_{\tL,w\}}^\gamma)\epsilon_{\alpha\beta\gamma}
+\cdots.
\end{align}
Here, the operator on the top belongs to the basis adopted in \cite{Liao:2020zyx},
while the first operator on the right-hand side belongs to our chosen basis in~\cref{eq:dim7basis}. 
Upon using the EoMs for the quark fields, the remaining two are EoM operators in the dim-6 BNV LEFT operator basis~\cite{Jenkins:2017jig}:
\begin{align}
&- 2 (\overline{l_\tR}q_{\tL,\{z}^\alpha)
(\overline{q_{\tL,w\}}^{\beta\C}}i\slashed{D} q_{\tR,y}^\gamma)\epsilon_{\alpha\beta\gamma}
\nonumber\\
&- 4 (\overline{l_\tR}q_{\tL,\{z}^\alpha)
(\overline{q_{\tR,y}^{\beta\C}}i\slashed{D} q_{\tL,w\}}^\gamma)\epsilon_{\alpha\beta\gamma}
\nonumber\\
\overset{\rm EoM}{=}&- 2 m_y (\overline{l_\tR} q_{\tL,\{z}^\alpha) (\overline{ q_{\tL, w\} }^{\beta \C} }  q_{\tL, y}^\gamma ) \epsilon_{\alpha \beta \gamma}
\nonumber\\
&- 2 m_z (\overline{l_\tR} q_{\tL,w}^\alpha) (\overline{q_{\tR,y}^{\beta\C}} q_{\tR,z}^\gamma) \epsilon_{\alpha \beta \gamma}
\nonumber\\
&- 2 m_w (\overline{l_\tR} q_{\tL,z}^\alpha) (\overline{q_{\tR,y}^{\beta\C}} q_{\tR,w}^\gamma) \epsilon_{\alpha \beta \gamma}.
\end{align}

\section{Summary of spurion fields}
\label{app:spurions}

The explicit expressions of the spurion fields $\calP_{xyz}^{\tL\tR,\mu}$ and 
$\calP_{xyz}^{\tL\tL,\mu}$ associated with the dim-7 BNV LEFT operators are: 
\begin{subequations}
\label{eq:spurions}
\begin{align}  
\calP_{uuu}^{\tL\tR,\mu} =\,&0, &
\calP_{uud}^{\tL\tR,\mu} =\,& 
C_{\partial\ell uud}^{\tL\tR}
i\overline{D^\mu\ell_\tL^\C}, 
\nonumber\\
\calP_{udu}^{\tL\tR,\mu} =\,&
\frac{1}{2} C_{\partial\ell udu}^{\tL\tR}
i\overline{D^\mu\ell_\tL^\C}, &
\calP_{uus}^{\tL\tR,\mu} =\,& 
C_{\partial\ell uus}^{\tL\tR}
i\overline{D^\mu\ell_\tL^\C}, 
\nonumber\\
\calP_{usu}^{\tL\tR,\mu} =\,&
\frac{1}{2} C_{\partial\ell usu}^{\tL\tR}
i\overline{D^\mu\ell_\tL^\C}, &
\calP_{ddd}^{\tL\tR,\mu} =\,&
C_{\partial\bar\ell ddd}^{\tL\tR}
i\overline{D^\mu\ell_\tR}, 
\nonumber\\
\calP_{dds}^{\tL\tR,\mu} =\,&
C_{\partial\bar\ell dds}^{\tL\tR}
i\overline{D^\mu\ell_\tR}, &
\calP_{dsd}^{\tL\tR,\mu} =\,&
\frac{1}{2}C_{\partial\bar\ell dsd}^{\tL\tR}
i\overline{D^\mu\ell_\tR},
\nonumber\\
\calP_{ddu}^{\tL\tR,\mu} =\,&
C_{\partial\nu ddu}^{\tL\tR}
i\overline{\partial^\mu\nu_\tL^\C}, & 
\calP_{udd}^{\tL\tR,\mu} =\,&
\frac{1}{2}C_{\partial\nu udd}^{\tL\tR}
i\overline{\partial^\mu\nu_\tL^\C},
\nonumber\\
\calP_{dsu}^{\tL\tR,\mu} =\,&
\frac{1}{2}C_{\partial\nu dsu}^{\tL\tR}
i\overline{\partial^\mu\nu_\tL^\C}, &
\calP_{usd}^{\tL\tR,\mu} =\,&
\frac{1}{2}C_{\partial\nu usd}^{\tL\tR}
i\overline{\partial^\mu\nu_\tL^\C}
\nonumber\\
\calP_{uds}^{\tL\tR,\mu} =\,&
\frac{1}{2}C_{\partial\nu uds}^{\tL\tR}
i\overline{\partial^\mu\nu_\tL^\C}, & 
\calP_{uss}^{\tL\tR,\mu} =\,&
\frac{1}{2}C_{\partial\nu uss}^{\tL\tR}
i\overline{\partial^\mu\nu_\tL^\C}, 
\nonumber\\
\calP_{ssu}^{\tL\tR,\mu} =\,&
C_{\partial\nu ssu}^{\tL\tR}
i\overline{\partial^\mu\nu_\tL^\C}, &
\calP_{dss}^{\tL\tR,\mu} =\,&
\frac{1}{2}C_{\partial\bar\ell dss}^{\tL\tR}
i\overline{D^\mu\ell_\tR}, 
\nonumber\\
\calP_{ssd}^{\tL\tR,\mu} =\,&
C_{\partial\bar\ell ssd}^{\tL\tR}
i\overline{D^\mu\ell_\tR}, &
\calP_{sss}^{\tL\tR,\mu} =\,&
C_{\partial\bar\ell sss}^{\tL\tR}
i\overline{D^\mu\ell_\tR}, 
\\%
\calP_{uuu}^{\tL\tL,\mu}=\,&0, &
\calP_{uud}^{\tL\tL,\mu}=\,&
\frac{1}{3}C_{\partial\ell duu}^{\tL\tL}
\overline{\ell_\tR^\C}\gamma^\mu, 
\nonumber\\
\calP_{uus}^{\tL\tL,\mu}=\,&
\frac{1}{3}C_{\partial\ell suu}^{\tL\tL}
\overline{\ell_\tR^\C}\gamma^\mu, &
\calP_{ddd}^{\tL\tL,\mu}=\,&
C_{\partial\bar\ell ddd}^{\tL\tL}
\overline{\ell_\tL}\gamma^\mu,
\nonumber\\
\calP_{dds}^{\tL\tL,\mu}=\,&
\frac{1}{3}C_{\partial\bar\ell dds}^{\tL\tL}
\overline{\ell_\tL}\gamma^\mu, &
\calP_{udd}^{\tL\tL,\mu}=\,&
\frac{1}{3}C_{\partial\bar\nu udd}^{\tL\tL}
\overline{\nu_\tL}\gamma^\mu,
\nonumber\\
\calP_{uds}^{\tL\tL,\mu}=\,&
\frac{1}{6}C_{\partial\bar\nu uds}^{\tL\tL}
\overline{\nu_\tL}\gamma^\mu, &
\calP_{uss}^{\tL\tL,\mu}=\,&
\frac{1}{3}C_{\partial\bar\nu uss}^{\tL\tL}
\overline{\nu_\tL}\gamma^\mu,
\nonumber\\
\calP_{dss}^{\tL\tL,\mu}=\,&
\frac{1}{3}C_{\partial\bar\ell dss}^{\tL\tL}
\overline{\ell_\tL}\gamma^\mu, &
\calP_{sss}^{\tL\tL,\mu}=\,&
C_{\partial\bar\ell sss}^{\tL\tL}
\overline{\ell_\tL}\gamma^\mu. 
\end{align}
\end{subequations}
Their corresponding chirality-flipped partners are obtained with simultaneously interchanging $\tL\leftrightarrow \tR$ and $\nu_\tL\leftrightarrow \nu_\tL^\C$. 
Note that the spurion fields $\calP_{xyz}^{\tL\tR,\mu}$ and 
$\calP_{xyz}^{\tL\tL,\mu}$
inherit the same flavor symmetries as their triple-quark operators
$\calN_{xyz}^{\tL\tR,\mu}$ and 
$\calN_{xyz}^{\tL\tL,\mu}$, respectively.
Therefore, only the independent flavor combinations are listed in \cref{eq:spurions}.
The numerical factors arise from the difference in conventions between the LEFT operators and \cref{eq:Lag_lqqqD}.

\bibliography{references_paper}{}

@article{JUNO:2015zny,
    author = "An, Fengpeng and others",
    collaboration = "JUNO",
    title = "{Neutrino Physics with JUNO}",
    eprint = "1507.05613",
    archivePrefix = "arXiv",
    primaryClass = "physics.ins-det",
    doi = "10.1088/0954-3899/43/3/030401",
    journal = "J. Phys. G",
    volume = "43",
    number = "3",
    pages = "030401",
    year = "2016"
}

@article{Hyper-Kamiokande:2018ofw,
    author = "Abe, K. and others",
    collaboration = "Hyper-Kamiokande",
    title = "{Hyper-Kamiokande Design Report}",
    eprint = "1805.04163",
    archivePrefix = "arXiv",
    primaryClass = "physics.ins-det",
    month = "5",
    year = "2018"
}

@article{DUNE:2020ypp,
    author = "Abi, Babak and others",
    collaboration = "DUNE",
    title = "{Deep Underground Neutrino Experiment (DUNE), Far Detector Technical Design Report, Volume II: DUNE Physics}",
    eprint = "2002.03005",
    archivePrefix = "arXiv",
    primaryClass = "hep-ex",
    reportNumber = "FERMILAB-PUB-20-025-ND, FERMILAB-DESIGN-2020-02",
    month = "2",
    year = "2020"
}

@article{Theia:2019non,
    author = "Askins, M. and others",
    collaboration = "Theia",
    title = "{THEIA: an advanced optical neutrino detector}",
    eprint = "1911.03501",
    archivePrefix = "arXiv",
    primaryClass = "physics.ins-det",
    doi = "10.1140/epjc/s10052-020-7977-8",
    journal = "Eur. Phys. J. C",
    volume = "80",
    number = "5",
    pages = "416",
    year = "2020"
}

@article{Super-Kamiokande:2025lxa,
    author = "Jung, S. and others",
    collaboration = "Super-Kamiokande",
    title = "{Search for nucleon decay via p{\textrightarrow}{\ensuremath{\nu}}{\ensuremath{\pi}}+ and n{\textrightarrow}{\ensuremath{\nu}}{\ensuremath{\pi}}0 in 0.484 Mton-year of Super-Kamiokande data}",
    eprint = "2510.26232",
    archivePrefix = "arXiv",
    primaryClass = "hep-ex",
    doi = "10.1103/gwc6-55bg",
    journal = "Phys. Rev. D",
    volume = "113",
    number = "1",
    pages = "012015",
    year = "2026"
}

@article{Super-Kamiokande:2026yep,
    author = "Abe, K. and others",
    collaboration = "Super-Kamiokande",
    title = "{Search for proton decay via $p \to e^{+}\pi^{0}\pi^{0}$ and $p \to \mu^{+}\pi^{0}\pi^{0}$ in 0.401 megaton-years exposure of Super-Kamiokande I-V}",
    eprint = "2604.10975",
    archivePrefix = "arXiv",
    primaryClass = "hep-ex",
    month = "4",
    year = "2026"
}

@article{Kamiokande:2026net,
    author = "Liu, Y. M. and others",
    collaboration = "Kamiokande",
    title = "{Search for proton decay into a single charged antilepton and a massless invisible particle using the full pure water data set of Super-Kamiokande}",
    eprint = "2608.30361",
    archivePrefix = "arXiv",
    primaryClass = "hep-ex",
    month = "8",
    year = "2026"
}

@article{JUNO:2025gmd,
    author = "Abusleme, Angel and others",
    collaboration = "JUNO",
    title = "{Measurement of reactor neutrino oscillation with the first JUNO data}",
    eprint = "2511.14593",
    archivePrefix = "arXiv",
    primaryClass = "hep-ex",
    doi = "10.1038/s41586-026-10538-z",
    journal = "Nature",
    volume = "654",
    number = "8118",
    pages = "343--348",
    year = "2026"
}

@article{JUNO:2025fpc,
    author = "Abusleme, Angel and others",
    collaboration = "JUNO",
    title = "{Initial performance results of the JUNO detector*}",
    eprint = "2511.14590",
    archivePrefix = "arXiv",
    primaryClass = "hep-ex",
    doi = "10.1088/1674-1137/ae3dc1",
    journal = "Chin. Phys. C",
    volume = "50",
    number = "4",
    pages = "043001",
    year = "2026"
}

@article{Jenkins:2017jig,
    author = "Jenkins, Elizabeth E. and Manohar, Aneesh V. and Stoffer, Peter",
    title = "{Low-Energy Effective Field Theory below the Electroweak Scale: Operators and Matching}",
    eprint = "1709.04486",
    archivePrefix = "arXiv",
    primaryClass = "hep-ph",
    doi = "10.1007/JHEP03(2018)016",
    journal = "JHEP",
    volume = "03",
    pages = "016",
    year = "2018",
    note = "[Erratum: JHEP 12, 043 (2023)]"
}

@article{Liao:2020zyx,
    author = "Liao, Yi and Ma, Xiao-Dong and Wang, Quan-Yu",
    title = "{Extending low energy effective field theory with a complete set of dimension-7 operators}",
    eprint = "2005.08013",
    archivePrefix = "arXiv",
    primaryClass = "hep-ph",
    doi = "10.1007/JHEP08(2020)162",
    journal = "JHEP",
    volume = "08",
    pages = "162",
    year = "2020"
}

@article{Jenkins:1990jv,
    author = "Jenkins, Elizabeth Ellen and Manohar, Aneesh V.",
    title = "{Baryon chiral perturbation theory using a heavy fermion Lagrangian}",
    reportNumber = "UCSD-PTH-90-23",
    doi = "10.1016/0370-2693(91)90266-S",
    journal = "Phys. Lett. B",
    volume = "255",
    pages = "558--562",
    year = "1991"
}

@article{Bijnens:1985kj,
    author = "Bijnens, J. and Sonoda, H. and Wise, Mark B.",
    title = "{On the Validity of Chiral Perturbation Theory for Weak Hyperon Decays}",
    reportNumber = "CALT-68-1221",
    doi = "10.1016/0550-3213(85)90569-3",
    journal = "Nucl. Phys. B",
    volume = "261",
    pages = "185--198",
    year = "1985"
}

@article{Claudson:1981gh,
    author = "Claudson, Mark and Wise, Mark B. and Hall, Lawrence J.",
    title = "{Chiral Lagrangian for Deep Mine Physics}",
    reportNumber = "HUTP-81/A036",
    doi = "10.1016/0550-3213(82)90401-1",
    journal = "Nucl. Phys. B",
    volume = "195",
    pages = "297--307",
    year = "1982"
}

@article{Liao:2025vlj,
    author = "Liao, Yi and Ma, Xiao-Dong and Wang, Hao-Lin",
    title = "{New Chiral Structures for Baryon Number Violating Nucleon Decays}",
    eprint = "2504.14855",
    archivePrefix = "arXiv",
    primaryClass = "hep-ph",
    doi = "10.1103/d8m7-5xxx",
    journal = "Phys. Rev. Lett.",
    volume = "135",
    number = "16",
    pages = "161801",
    year = "2025"
}

@article{Liao:2025sqt,
    author = "Liao, Yi and Ma, Xiao-Dong and Wang, Hao-Lin",
    title = "{Chiral perturbation theory for baryon-number-violating nucleon decay into a vector meson}",
    eprint = "2506.05052",
    archivePrefix = "arXiv",
    primaryClass = "hep-ph",
    doi = "10.1103/fzq9-tfcp",
    journal = "Phys. Rev. D",
    volume = "112",
    number = "3",
    pages = "L031704",
    year = "2025"
}

@article{Beneito:2023xbk,
    author = "Beneito, I, Arnau Bas and Gargalionis, John and Herrero-Garcia, Juan and Santamaria, Arcadi and Schmidt, Michael A.",
    title = "{An EFT approach to baryon number violation: lower limits on the new physics scale and correlations between nucleon decay modes}",
    eprint = "2312.13361",
    archivePrefix = "arXiv",
    primaryClass = "hep-ph",
    reportNumber = "IFIC/23-52, CPPC-2023-12",
    doi = "10.1007/JHEP07(2024)004",
    journal = "JHEP",
    volume = "07",
    pages = "004",
    year = "2024",
    note = "[Erratum: JHEP 02, 065 (2026)]"
}

@article{Gargalionis:2024nij,
    author = "Gargalionis, John and Herrero-Garc{\'\i}a, Juan and Schmidt, Michael A.",
    title = "{Model-independent estimates for loop-induced baryon-number-violating nucleon decays}",
    eprint = "2401.04768",
    archivePrefix = "arXiv",
    primaryClass = "hep-ph",
    reportNumber = "CPPC-2024-02",
    doi = "10.1007/JHEP06(2024)182",
    journal = "JHEP",
    volume = "06",
    pages = "182",
    year = "2024"
}

@article{Chen:2025mjt,
    author = "Chen, Jing and Liao, Yi and Ma, Xiao-Dong and Wang, Hao-Lin",
    title = "{Nucleon decays into three leptons: Noncontact contributions}",
    eprint = "2512.02692",
    archivePrefix = "arXiv",
    primaryClass = "hep-ph",
    doi = "10.1016/j.physletb.2026.140401",
    journal = "Phys. Lett. B",
    volume = "876",
    pages = "140401",
    year = "2026"
}

@article{Liao:2025wxk,
    author = "Liao, Yi and Ma, Xiao-Dong and Zhao, Xiang",
    title = "{Nucleon decays into three leptons: contact contributions}",
    eprint = "2512.09287",
    archivePrefix = "arXiv",
    primaryClass = "hep-ph",
    doi = "10.1007/JHEP05(2026)151",
    journal = "JHEP",
    volume = "05",
    pages = "151",
    year = "2026"
}

@article{Fan:2026fqo,
    author = "Fan, Wei-Qi and Liao, Yi and Ma, Xiao-Dong",
    title = "{Nucleon decays into one lepton plus two non-strange mesons}",
    eprint = "2604.24613",
    archivePrefix = "arXiv",
    primaryClass = "hep-ph",
    month = "4",
    year = "2026"
}

@article{Fan:2026csl,
    author = "Fan, Wei-Qi and Liao, Yi and Ma, Xiao-Dong",
    title = "{Comprehensive investigation of nucleon decays into one lepton plus two mesons}",
    eprint = "2605.17525",
    archivePrefix = "arXiv",
    primaryClass = "hep-ph",
    month = "5",
    year = "2026"
}

@article{Liao:2016hru,
    author = "Liao, Yi and Ma, Xiao-Dong",
    title = "{Renormalization Group Evolution of Dimension-seven Baryon- and Lepton-number-violating Operators}",
    eprint = "1607.07309",
    archivePrefix = "arXiv",
    primaryClass = "hep-ph",
    doi = "10.1007/JHEP11(2016)043",
    journal = "JHEP",
    volume = "11",
    pages = "043",
    year = "2016"
}

@article{Lehman:2014jma,
    author = "Lehman, Landon",
    title = "{Extending the Standard Model Effective Field Theory with the Complete Set of Dimension-7 Operators}",
    eprint = "1410.4193",
    archivePrefix = "arXiv",
    primaryClass = "hep-ph",
    doi = "10.1103/PhysRevD.90.125023",
    journal = "Phys. Rev. D",
    volume = "90",
    number = "12",
    pages = "125023",
    year = "2014"
}

@article{Seidel:1988ut,
    author = "Seidel, S. and others",
    title = "{Search for Multitrack Nucleon Decay}",
    doi = "10.1103/PhysRevLett.61.2522",
    journal = "Phys. Rev. Lett.",
    volume = "61",
    pages = "2522--2525",
    year = "1988"
}

@article{ParticleDataGroup:2024cfk,
    author = "Navas, S. and others",
    collaboration = "Particle Data Group",
    title = "{Review of particle physics}",
    doi = "10.1103/PhysRevD.110.030001",
    journal = "Phys. Rev. D",
    volume = "110",
    number = "3",
    pages = "030001",
    year = "2024"
}

@article{Weinberg:1989dx,
	author = "Weinberg, Steven",
	title = "{Larger Higgs Exchange Terms in the Neutron Electric Dipole Moment}",
	reportNumber = "UTTG-30-89",
	doi = "10.1103/PhysRevLett.63.2333",
	journal = "Phys. Rev. Lett.",
	volume = "63",
	pages = "2333",
	year = "1989"
}

@article{Song:2026gyo,
    author = "Song, Chuan-Qiang and Yu, Jiang-Hao",
    title = "{Comprehensive Effective Field Theory Analysis for Baryon Number Violating Processes}",
    eprint = "2603.11158",
    archivePrefix = "arXiv",
    primaryClass = "hep-ph",
    month = "3",
    year = "2026"
}

@article{Bali:2022qja,
    author = {Bali, Gunnar S. and Collins, Sara and S\"oldner, Wolfgang and Weish\"aupl, Simon},
    collaboration = "RQCD",
    title = "{Leading order mesonic and baryonic SU(3) low energy constants from Nf=3 lattice QCD}",
    eprint = "2201.05591",
    archivePrefix = "arXiv",
    primaryClass = "hep-lat",
    doi = "10.1103/PhysRevD.105.054516",
    journal = "Phys. Rev. D",
    volume = "105",
    number = "5",
    pages = "054516",
    year = "2022"
}

@article{Liao:2026ugl,
    author = "Liao, Yi and Ma, Xiao-Dong and Zhao, Xiang",
    title = "{Renormalization-group-improved constraints on dimension-7 baryon-number-violating operators}",
    eprint = "2604.00952",
    archivePrefix = "arXiv",
    primaryClass = "hep-ph",
    doi = "10.1016/j.physletb.2026.140678",
    journal = "Phys. Lett. B",
    volume = "879",
    pages = "140678",
    year = "2026"
}

@article{Frejus:1991ben,
    author = "Berger, Christoph and others",
    collaboration = "Frejus",
    title = "{Lifetime limits on (B-L) violating nucleon decay and dinucleon decay modes from the Frejus experiment}",
    reportNumber = "WUB-91-21",
    doi = "10.1016/0370-2693(91)91479-F",
    journal = "Phys. Lett. B",
    volume = "269",
    pages = "227--233",
    year = "1991"
}

@article{Fuentes-Martin:2022jrf,
    author = {Fuentes-Mart{\'\i}n, Javier and K{\"o}nig, Matthias and Pag{\`e}s, Julie and Thomsen, Anders Eller and Wilsch, Felix},
    title = "{A proof of concept for matchete: an automated tool for matching effective theories}",
    eprint = "2212.04510",
    archivePrefix = "arXiv",
    primaryClass = "hep-ph",
    reportNumber = "MITP-22-105, TUM-HEP-1443/22, ZU-TH-58/22",
    doi = "10.1140/epjc/s10052-023-11726-1",
    journal = "Eur. Phys. J. C",
    volume = "83",
    number = "7",
    pages = "662",
    year = "2023"
}
\bibliographystyle{utphys}

\end{document}